\documentclass[11pt]{article}

\usepackage[top=0.9in, bottom=0.9in, left=0.9in, right=0.9in]{geometry}

\usepackage{microtype}
\usepackage{lmodern}
\usepackage[T1]{fontenc}
\usepackage[utf8]{inputenc}

\usepackage{amsmath}
\usepackage{amssymb}

\usepackage{booktabs}
\usepackage{graphicx}
\usepackage{multirow}
\usepackage{float}
\usepackage{wrapfig}
\usepackage{tabularx}

\usepackage{enumitem}

\usepackage{xcolor}
\definecolor{darkblue}{rgb}{0, 0, 0.5}
\usepackage{hyperref}
\hypersetup{colorlinks=true, citecolor=darkblue, linkcolor=darkblue, urlcolor=darkblue}

\usepackage{fancyhdr}
\renewcommand{\headrulewidth}{0pt}
\fancypagestyle{firstpage}{%
  \fancyhf{}%
  \renewcommand{\headrulewidth}{0pt}%
  \fancyhead[R]{}%
  \fancyfoot[L]{\small\textit{Corresponding author: Yiheng Yao $\langle$\href{mailto:yaoyh@stanford.edu}{yaoyh@stanford.edu}$\rangle$}}%
  \fancyfoot[R]{\thepage}%
}

\usepackage{titlesec}
\titleformat{\section}{\large\bfseries}{{\normalfont\large\bfseries\thesection.}}{0.5em}{}
\titleformat{\subsection}{\normalsize\bfseries}{{\normalfont\normalsize\bfseries\thesubsection}}{0.5em}{}
\titleformat{\paragraph}[runin]{\normalsize\bfseries}{}{0em}{}[.\quad]
\titlespacing*{\section}{0pt}{1.4ex plus .3ex minus .2ex}{0.8ex plus .1ex}
\titlespacing*{\subsection}{0pt}{1.2ex plus .2ex minus .1ex}{0.6ex}

\usepackage{abstract}

\usepackage[labelfont=bf, font=small]{caption}

\usepackage{natbib}

\usepackage[title,titletoc]{appendix}

\usepackage{url}

\usepackage{fvextra}
\fvset{breaklines=true, breakanywhere=true, breaksymbol={}}
\AtBeginDocument{%
  \DefineVerbatimEnvironment{verbatim}{Verbatim}{breaklines=true,breakanywhere=true,breaksymbol={}}%
}

\usepackage[most]{tcolorbox}
\definecolor{cardinal}{rgb}{0.60, 0.067, 0.106}  
\definecolor{cardinalbg}{rgb}{0.98, 0.93, 0.93}  

\tcbset{
  herofig/.style={
    enhanced,
    colback=cardinalbg,
    colframe=cardinal,
    arc=8pt,
    boxrule=1pt,
    left=10pt, right=10pt, top=10pt, bottom=8pt,
    drop shadow={cardinal!30!black, opacity=0.25},
  }
}

\begin{document}
\thispagestyle{firstpage}
\setlength{\parskip}{0.5ex}

\noindent{\LARGE\bfseries Talk is Cheap, Communication is Hard: Dynamic Grounding Failures and Repair in Multi-Agent Negotiation}

\vspace{1.2em}
\noindent{\normalsize Yiheng Yao, Chelsea Zou, and Robert D. Hawkins}

\vspace{0.3em}
\noindent{\small Stanford University}

\vspace{0.4em}
\noindent{\small \textit{Preprint. Under review.}}

\vspace{1em}

\begin{abstract}
\noindent Grounding is the collaborative process of establishing mutual belief sufficient for a communicative goal. While static grounding maps language to a shared context, dynamic grounding requires agents to negotiate meaning across turns. Current multi-agent Large Language Model (LLM) benchmarks largely emphasize static, one-shot tasks, overlooking whether agents can repair grounding breakdowns through interaction. We introduce an iterated multi-turn negotiation game where two agents allocate shared resources to private projects with verifiable jointly optimal outcomes. Although individual agents can identify Pareto-optimal allocations in isolation, agent dyads consistently fail to reach them across models. We identify four failure modes: (1) loss of shared interaction history, (2) stubborn anchoring to early proposals, (3) defaulting to equal splits over reward-maximizing coordination, and (4) referential binding errors across turns. Our baselines show that the coordination gap is not explained by individual reasoning limits or insufficient information exchange alone. Instead, the bottleneck lies in dynamic grounding: joint plan formation, commitment, and execution.
\end{abstract}

\vspace{0.8em}
\noindent\textit{Keywords: multi-agent communication, grounding, negotiation, resource allocation, discourse pragmatics}

\vspace{1em}
\hrule
\vspace{1em}

\section{Introduction}
\label{sec:intro}

Successful communication requires more than exchanging propositions; it requires \emph{grounding}---the collaborative process by which interlocutors establish that their utterances have been understood well enough for current purposes \citep{clark1991grounding}. In human dialogue, grounding is an active, incremental process: speakers monitor comprehension, request clarification, and repair misunderstandings in real time. This \emph{dynamic} grounding stands in contrast to \emph{static} grounding, which links language to a fixed, externally observable context such as a knowledge base or image \citep{chandu2021grounding, anikina2025common}. Dynamic grounding is central to real-world coordination problems where agents must allocate shared resources under partial information. For example, project managers coordinating engineering hours, compute, and testing bandwidth across concurrent workstreams must explain priorities, negotiate trade-offs, and revise plans over repeated meetings. As AI systems increasingly act on users' behalf, these coordination demands become questions of faithful representation: an assistant must preserve a user's goals, constraints, and commitments while negotiating shared plans with other agents.

Current multi-agent LLM benchmarks largely focus on static tasks: matching descriptions to images \citep{eisenstein2026mtpingeval}, optimizing fixed payoff matrices \citep{bianchi2024negotiationarena}, or agreeing on predefined prices \citep{davidson2024evaluating}. These settings test whether models can exchange facts, but not whether they can \emph{build and maintain shared understanding through interaction}, the hallmark of dynamic grounding.

\begin{figure}[t]
  \begin{tcolorbox}[herofig]
    \centering
    \includegraphics[width=\linewidth]{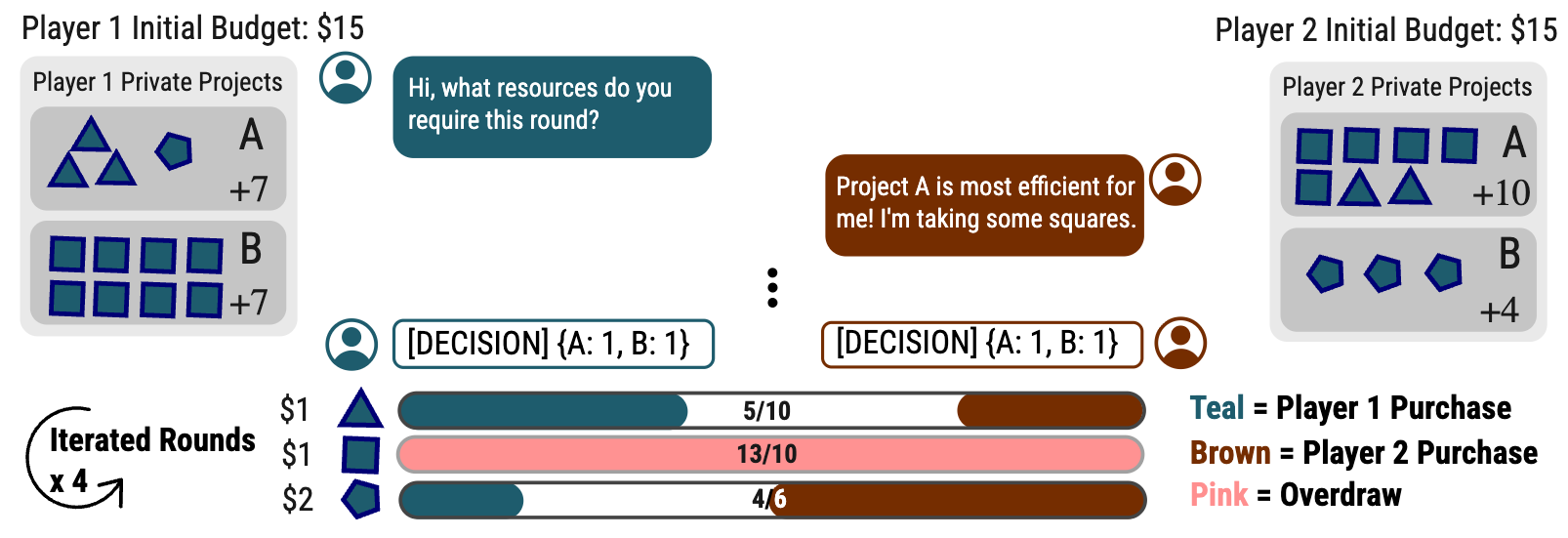}
    \captionof{figure}{Illustration of the resource allocation game agents play. Each agent has a private set of projects with different requirements and makes purchases from a common pool of resources after exchanging up to 5 messages each. If joint resource purchases exceed capacity, an overdraw occurs and no rewards are given to either agent. The game is iterated over 4 rounds with the same or different partner and projects. See \S\ref{sec:game:conditions} for a breakdown of the different conditions studied.}
    \label{fig:game_illustration}
  \end{tcolorbox}
\end{figure}

We argue that the gap between static and dynamic grounding is a key bottleneck for multi-agent LLM coordination. To study this, we introduce an \emph{iterated, multi-turn} negotiation game in which two agents share a pool of resources and must allocate them toward private \emph{projects} with different resource requirements and rewards (Figure~\ref{fig:game_illustration}). Within each round, agents engage in multi-turn cheap talk before making simultaneous purchase decisions; across rounds, they face the same or new partners and scenarios, enabling the study of grounding accumulation, repair, and adaptation over repeated encounters. The asymmetric project structure forces agents to explain not only \emph{what} they want, but \emph{why}: which projects they pursue, which resources are flexible, and how to avoid costly overdraw penalties. Because scenarios have verifiable joint optima, we can measure the gap between what agents \emph{could} achieve and what they actually achieve, while using interaction traces to diagnose where grounding breaks down.

Our experiments show that individual agents can identify optimal strategies in isolation, but interacting dyads consistently underperform. We identify four failure modes: missing shared interaction history, stubborn anchoring to early proposals, defaulting to equal splits over reward-maximizing allocations, and referential binding failures where agents lose track of commitments across turns. Our contributions are: (i)~a configurable negotiation game environment that isolates the costs of interactive coordination from individual reasoning capacity, with verifiable outcomes; (ii)~an empirical decomposition of the coordination gap showing that the bottleneck lies in interactive plan formation and commitment, not reasoning limitations or information asymmetry, together with a characterization of four failure modes; and (iii)~a framework for studying interventions including better prompt design, structured agent interaction patterns, and collecting game traces for offline training in the future.

\section{Related work}
\label{sec:related}

Our work sits at the intersection of three lines of research: grounding in dialogue, LLM-based negotiation, and multi-agent coordination. We review each and identify the gap our framework addresses.

\subsection{Grounding in dialogue}
\label{sec:related:grounding}

Grounding, introduced by \citet{clark1991grounding}, is the process of establishing mutual belief sufficient for current purposes. In cognitive science, it includes both static symbol grounding, mapping language to perceptual or external referents, and dynamic communicative grounding, where shared understanding is built interactively over mutually presupposed common ground \citep{stalnaker2002common}. NLP has largely emphasized the former while neglecting the latter \citep{chandu2021grounding}; a recent survey of 448 papers similarly finds that evaluation frameworks for dynamic grounding in Language Model (LM) interactions remain scarce \citep{anikina2025common}. We target this gap with a task that requires agents to communicate, negotiate shared plans, repair misunderstandings, and track commitments across turns to achieve optimal outcomes.

\subsection{LLM-based negotiation}
\label{sec:related:negotiation}

Recent work evaluates LLMs in structured negotiation settings, including landlord--tenant bargaining \citep{davidson2024evaluating}, six-agent multi-issue negotiation \citep{abdelnabi2024negotiation}, resource exchange and ultimatum games \citep{bianchi2024negotiationarena}, equilibrium computation \citep{hua2024game}, Deal-or-No-Deal self-play \citep{liao2024efficacy}, and multi-party bargaining with hidden rewards \citep{qian2026strategic}. Across these settings, two gaps remain. First, negotiations are often \emph{multi-turn} but not \emph{iterated}: agents exchange messages before one decision, with little chance to repair grounding across repeated encounters. Second, evaluation emphasizes \emph{outcome metrics}, such as deal rates, scores, and Pareto efficiency, rather than the communicative \emph{process}. Table~\ref{tab:comparison} summarizes how our framework addresses both gaps by combining multi-turn cheap talk with simultaneous decisions across iterated rounds with stable or shifting partners, enabling analysis of grounding accumulation and repair.

\begin{table}[t]
\centering
\small
\begin{tabular}{lccccc}
\toprule
\textbf{Framework} & \textbf{Multi-turn} & \textbf{Iterated} & \textbf{Cheap talk} & \textbf{Private info} & \textbf{Eval.\ focus} \\
\midrule
\citet{davidson2024evaluating}     & \checkmark & ---        & \checkmark & \checkmark & Outcome \\
\citet{abdelnabi2024negotiation}   & \checkmark & ---        & \checkmark & \checkmark & Outcome \\
\citet{bianchi2024negotiationarena}& \checkmark & ---        & \checkmark & \checkmark & Outcome \\
\citet{eisenstein2026mtpingeval}   & \checkmark & ---        & \checkmark & \checkmark & Process \\
\citet{qian2026strategic}          & ---        & \checkmark & ---        & \checkmark & Outcome \\
\citet{madmoun2026communication}   & ---        & \checkmark & Minimal    & ---        & Outcome \\
\textbf{Ours}                      & \checkmark & \checkmark & \checkmark & \checkmark & Both \\
\bottomrule
\end{tabular}
\caption{Comparison of multi-agent LLM evaluation frameworks. \emph{Multi-turn}: multiple exchanges within a single decision round. \emph{Iterated}: repeated decision rounds with the same or different partners. \emph{Private info}: agents hold hidden reward-relevant information. \emph{Eval focus}: whether analysis targets negotiation outcomes or the communicative process.}
\label{tab:comparison}
\end{table}

\subsection{Multi-agent coordination}
\label{sec:related:cooperation}

MT-PingEval \citep{eisenstein2026mtpingeval} evaluates collaborative private-information games under fixed token budgets and variable turn counts, introducing a hierarchy of \emph{interactivity levels} (see Appendix~\ref{app:interactivity} for where our game falls). It finds that LLMs often fail to benefit from additional turns, producing content-rich utterances that they ``struggle to deploy strategically in service of collaborative goals.'' We extend this paradigm to an iterated setting, testing whether agents can recover from such failures across repeated encounters. The hidden profile paradigm \citep{stasser1985pooling}, where optimal group decisions require pooling private information, has recently been applied to LLM agents in HiddenBench \citep{li2026hiddenbench}. This reveals a large gap between single agents with complete information and multi-agent groups that must pool asymmetric evidence through discussion. Our game adds both compatible and conflicting goals, with information asymmetry emerging from private project assignments. Related work shows that even minimal communication, such as a single-word channel, can increase cooperation from 0\% to 48.3\% in Stag Hunt \citep{madmoun2026communication}, though such channels cannot support the richer referential grounding our task requires. Most recently, \citet{khatua2026cooperbench} find coordination failures in collaborative coding teams, including vague messages, commitment violations, and incorrect partner expectations, with dyads performing 30\% below solo baselines. We isolate these communicative failures in a controlled setting, enabling systematic analysis of when and why coordination breaks down.

\section{Game environment}
\label{sec:game}

Two agents share a pool of resources (e.g., wood, stone, gold), each with a fixed per-round supply and unit cost. Each agent has an independent budget and simultaneously submits a purchase decision each round; if combined demand for any resource exceeds supply, the round is \emph{annulled} and both agents receive zero reward. Rather than assigning scalar values to resources directly, each agent is assigned private \emph{projects}: combinatorial goals requiring specific resource mixes and yielding rewards per completed run (e.g., 3~wood + 2~stone $\rightarrow$ 50~points). Unspent budget has no value. This project layer shifts the pragmatics from negotiating quantities to communicating \emph{why} particular combinations matter, creating a natural need for referential grounding around project names, requirements, and strategies. Using the interactivity hierarchy of \citet{eisenstein2026mtpingeval}, we show in Appendix~\ref{app:interactivity} that this structure makes the game at least level-3 interactive but not level-2 interactive; meaning that at least 3 messages need to be shared before the optimal joint allocation can be solved.

Each game consists of multiple rounds (set to 4 for our study). Within each round, agents first engage in \emph{cheap talk}: up to 5 turns each of alternating natural-language messages, during which they may share project information, propose allocations, or negotiate. Each agent also maintains a private \emph{thinking} scratchpad hidden from the opponent, enabling comparison of private intent with public speech. Agents then independently submit a resource allocation together with project assignments as structured JSON. After both decisions, the round resolves: overdraw is checked, rewards are computed, and both agents observe the opponent's purchases. Opponent \emph{rewards} are not disclosed unless voluntarily shared.

\subsection{Scenario generation}
\label{sec:game:scenarios}

We parameterize goal conflict via the \emph{compatibility ratio} $M/C$: the maximum achievable joint reward $M$ divided by the sum of individual maxima $C = V_1 + V_2$ (each agent optimizing independently). When $M/C = 1.0$, optimal strategies are fully compatible; when $M/C < 1.0$, individual optima conflict. We generate scenarios at three levels: $M/C \in \{0.5, 0.8, 1.0\}$.

Scenarios are generated via \emph{simulated annealing}. The generator initializes six random projects (three per agent), then iteratively perturbs requirements, rewards, and assignments to minimize a composite loss targeting the desired $M/C$ ratio ($\pm 0.05$). The loss jointly enforces: equal individual maxima ($V_1 = V_2$); swap fairness (the joint optimum can favor either agent equally); multiple joint-optimal solutions; and individual affordability of all projects. Thereafter, an oracle validation step further restricts the candidate pool to those that all models studied in subsequent games can solve under full information with a single attempt (pass@1) individually. This oracle baseline establishes that the coordination gap is not attributable to individual reasoning limitations---it isolates interaction itself as the locus of difficulty. The full list of scenarios is listed in Appendix~\ref{app:scenarios}.

\section{Experimental setup}
\label{sec:experiments}

We evaluate models spanning multiple providers: Claude~4.5 Sonnet (Anthropic), GPT-5~mini (OpenAI), and Qwen~3.5 Flash (via OpenRouter); see Appendix~\ref{app:models} for model configuration details. Prompt templates are provided in Appendix~\ref{app:prompts}. $N=10$ games are played for each of $3 \times 2 \times 2 = 12$ conditions, with all-to-all model cross-play (6 unique pairings) yielding a total of 720 game traces. Each game consists of 4 iterated rounds for a total of 2880 dyadic multi-turn interactions leading to a decision. All traces (cheap talk transcripts, thinking logs, allocations, rewards, and experiment metadata) are stored for reproducibility and released for further analysis; see Appendix~\ref{app:data_exploration}. To mitigate confounds, each configuration is played twice with first-speaker roles swapped, controlling for order effects, and abstract resource identifiers are replaced at runtime with a list of pre-sampled thematic names to amortize content biases (Appendix~\ref{app:scenarios}).

\subsection{Conditions and controls}
\label{sec:game:conditions}

\begin{enumerate}[leftmargin=*, itemsep=2pt]
  \item \textbf{Compatibility ratio} ($M/C \in \{0.5, 0.8, 1.0\}$): varies the degree of goal conflict. Agents start without knowledge of the other parties' projects and must communicate to determine whether their goals are aligned or opposed.
  \item \textbf{Partner stability}: \emph{stable} (same partner) vs.\ \emph{shifting} (one agent's context resets each round). This evaluates whether agents can leverage shared history to form ad-hoc conventions \citep{hawkins2020partners}, a well-studied phenomenon in humans yet often overlooked in LLMs.
  \item \textbf{Project rotation}: \emph{fixed} vs.\ \emph{rotating} (new scenarios each round). This holds the compatibility ratio constant, preserving the level of goal conflict, but eliminates a fixed environment and tests whether agents can develop transferable coordination strategies rather than memorizing prior successful allocations.
\end{enumerate}

\subsection{Metrics}
\label{sec:experiments:metrics}

We organize metrics into \emph{outcome metrics} quantifying coordination quality and \emph{process metrics} analyzing communicative strategies (Table~\ref{tab:metrics}).

\begin{table}[H]
\small
\begin{tabular*}{\textwidth}{@{\extracolsep{\fill}}llp{0.62\textwidth}}
\toprule
\textbf{Type} & \textbf{Metric} & \textbf{Definition} \\
\midrule
Outcome & Overdraw rate & Fraction of rounds where demand exceeds supply \\
Outcome & Allocation efficiency & Achieved joint reward / oracle optimum \\
Outcome & Optimum rate & Fraction of rounds achieving joint optimal strategy \\
\midrule
Process & Game strategy taxonomy & Payoff alternation, win-stay, and lose-shift across repeated rounds \\
Process & First-proposal deference & Degree to which responder's allocation matches first proposal \\
Process & Allocation anchoring & Consecutive-round allocation similarity (exact match rate) \\
Process & Stated-vs-actual coherence & Commitment to previously grounded plans \\
\bottomrule
\end{tabular*}
\caption{Outcome metrics quantify the coordination gap; process metrics decompose its behavioral sources.}
\label{tab:metrics}
\end{table}

\section{Results}
\label{sec:results}

Table~\ref{tab:crossplay} summarizes coordination quality across models, pairings, and conditions. To decompose the coordination gap, we anchor results against three baselines: an \emph{oracle baseline} (agents solving the allocation in isolation with full information) rules out individual reasoning limitations; a \emph{no-talk baseline} (\S\ref{sec:results:notalk}) establishes that communication is necessary; and a \emph{full-transparency intervention} (\S\ref{sec:results:transparency}) tests whether information asymmetry is the primary bottleneck. Within that frame, we isolate four behavioral failure modes (lack of shared interaction history, perfunctory fairness, anchoring, and referential binding failures) illustrated in Figure~\ref{fig:failure_modes}. Calibrated LLM-assisted trace annotations complement this picture in \S\ref{sec:llm-judge} (full taxonomy in Appendix~\ref{app:judge}).

\begin{table}[ht]
\centering
\caption{Self-play vs.\ cross-play performance. Overdraw (\%), Efficiency (\%), and Optimum (\%) rates are shown across compatibility ratios. Heterogeneous pairings (e.g., Sonnet 4.5 $\times$ GPT-5 Mini) frequently outperform self-play dyads.}
\label{tab:crossplay}
\small
\begin{tabular}{l ccc ccc ccc}
\toprule
& \multicolumn{3}{c}{Overdraw~$\downarrow$} & \multicolumn{3}{c}{Efficiency~$\uparrow$} & \multicolumn{3}{c}{Optimum~$\uparrow$} \\
\cmidrule(lr){2-4} \cmidrule(lr){5-7} \cmidrule(lr){8-10}
Pair & 0.5 & 0.8 & 1.0 & 0.5 & 0.8 & 1.0 & 0.5 & 0.8 & 1.0 \\
\midrule
GPT-5 Mini                     & 22.7 & \phantom{0}9.4 & \phantom{0}3.8 & 68.2 & 85.6 & 94.0 & 38.4 & 66.3 & 86.9 \\
Sonnet 4.5                     & 30.0 & 14.4 & \phantom{0}6.3 & 63.7 & 79.4 & 91.1 & 27.5 & 53.1 & 79.4 \\
Qwen 3.5                       & 38.8 & 32.5 & 16.3 & 56.8 & 61.2 & 76.1 & 32.5 & 43.8 & 43.8 \\
\addlinespace
Sonnet 4.5 $\times$ GPT-5 Mini & 10.6 & \phantom{0}5.0 & \phantom{0}1.3 & 79.8 & 89.7 & 97.9 & 34.4 & 66.3 & 91.3 \\
GPT-5 Mini $\times$ Qwen 3.5   & 16.9 & 13.8 & \phantom{0}6.3 & 74.0 & 77.4 & 87.4 & 31.9 & 57.5 & 69.4 \\
Qwen 3.5 $\times$ Sonnet 4.5   & 29.4 & 16.9 & \phantom{0}7.5 & 63.8 & 74.1 & 85.1 & 21.9 & 43.1 & 52.5 \\
\bottomrule
\end{tabular}
\end{table}

Across all conditions, 15.7\% of rounds result in overdraw, concentrated in competitive scenarios ($M/C{=}0.5$: 24.7\%) and rare in compatible ones ($M/C{=}1.0$: 6.9\%). Despite this, most dyads eventually find effective coordination: 81.7\% of games contain at least one jointly optimal round, and in the stable condition the first optimum is reached after an average of 1.6 rounds. After an overdraw, 83.3\% of subsequent rounds yield positive scores, suggesting that coordination failures serve as effective learning signals.

\begin{figure}[htbp]
  \centering
  \includegraphics[width=0.8\linewidth]{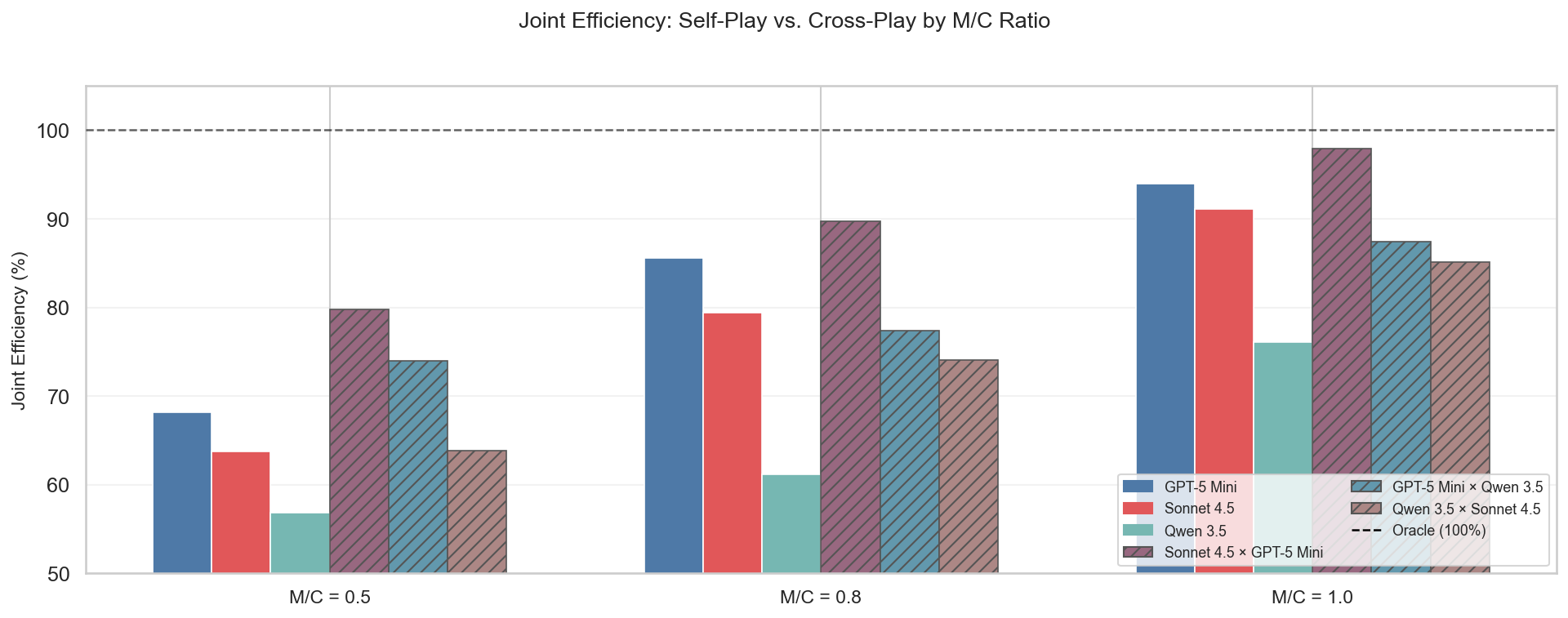}
  \caption{Joint efficiency for self-play (solid) and cross-play (hatched) dyads across compatibility ratios.}
  \label{fig:crossplay}
\end{figure}

Evaluating cross-play pairings reveals that heterogeneous dyads consistently outperform self-play under competitive conditions (Figure~\ref{fig:crossplay}). At $M/C{=}0.5$, \emph{every} cross-play pair exceeds the stronger model's self-play baseline: Sonnet~4.5 $\times$ GPT-5 Mini achieves 79.8\% joint efficiency (vs.\ 68.2\% and 63.7\% self-play), GPT-5 Mini $\times$ Qwen~3.5 reaches 74.0\% (vs.\ 68.2\%), and even Qwen~3.5 $\times$ Sonnet~4.5 matches Sonnet's self-play at 63.8\%. This cross-play advantage under competition is consistent with behavioral diversity aiding coordination when goal conflict is high, yielding the lowest overdraw rates in the Sonnet~4.5 $\times$ GPT-5 Mini pairing (1.3\% in compatible scenarios, 10.6\% in competitive ones). As compatibility increases ($M/C{=}0.8$ and $1.0$), the cross-play advantage narrows and weaker models begin to drag down stronger partners. GPT-5 Mini $\times$ Qwen~3.5 falls below GPT-5 Mini self-play at $M/C{=}0.8$ (77.4\% vs.\ 85.6\%), suggesting that under reduced conflict, individual capability matters more than behavioral diversity.

\subsection{The value of cheap talk}
\label{sec:results:notalk}

\begin{figure}[ht]
  \centering
  \includegraphics[width=\linewidth]{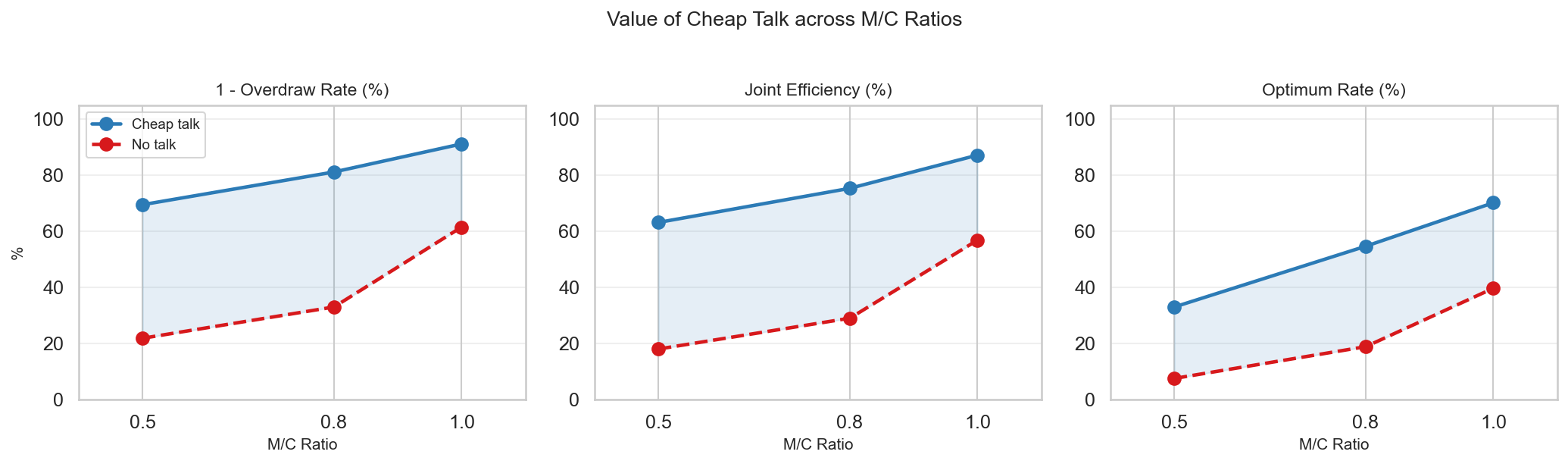}
  \caption{Value of cheap talk across compatibility ratios, aggregated over all models and conditions. Filled dots show cheap-talk performance; hollow dots show the no-talk baseline, and the shaded region represents the gain from communication. All three metrics are oriented so that higher is better: we report $1 - \text{overdraw rate}$ (fraction of rounds without supply violation), joint efficiency, and optimum rate. Cheap talk is most impactful under competitive conditions ($M/C{=}0.5$), where joint efficiency roughly triples (18.0\% $\rightarrow$ 63.2\%).}
  \label{fig:notalk}
\end{figure}

To quantify the contribution of communication, we compare dyads with multi-turn cheap talk against a \emph{no-talk baseline} where agents submit allocations without any communication (Figure~\ref{fig:notalk}; full numeric breakdown in Table~\ref{tab:notalk}, Appendix~\ref{app:results}). The baseline is run as self-play only ($N{=}4$ games per model $\times$ partner-stability $\times$ $M/C$ cell, yielding 32 rounds per cell).

Cheap talk produces large, consistent improvements across all models and conditions. Averaged across models and conditions, joint efficiency roughly triples under competition ($M/C{=}0.5$), rising from 18.0\% to 63.2\%. The gain persists under mixed ($M/C{=}0.8$: 29.0\% $\rightarrow$ 75.3\%) and fully compatible goals ($M/C{=}1.0$: 56.8\% $\rightarrow$ 87.2\%). Even naturally aligned agents benefit substantially from communication, since multiple allocation plans may satisfy both agents' individual optima and not all combinations are supply-compatible.

Supply violations tell a complementary story. Without communication, overdraw is pervasive at $M/C{=}0.5$: only 21.8\% of rounds avoid a supply violation, meaning roughly four in five competitive no-talk rounds end in failure. Cheap talk lifts this to 69.5\%, a 47.7-point gain. The gain remains large at $M/C{=}0.8$ (33.0\% $\rightarrow$ 81.2\%, a 48.2-point jump) and $M/C{=}1.0$ (61.5\% $\rightarrow$ 91.2\%, nearly 30 points). Optimum rates show the most dramatic lift: without communication, only 7.5\% of competitive rounds reach the joint maximum, while cheap talk raises this to 33.0\%; at $M/C{=}1.0$, optimum rates climb from 39.7\% to 70.2\%. These results confirm that in our experimental paradigm, cheap talk is indispensable to success.

\subsection{Lack of shared interaction history}
\label{sec:results:procedural}

\begin{figure}[ht]
  \centering
  \includegraphics[width=0.8\linewidth]{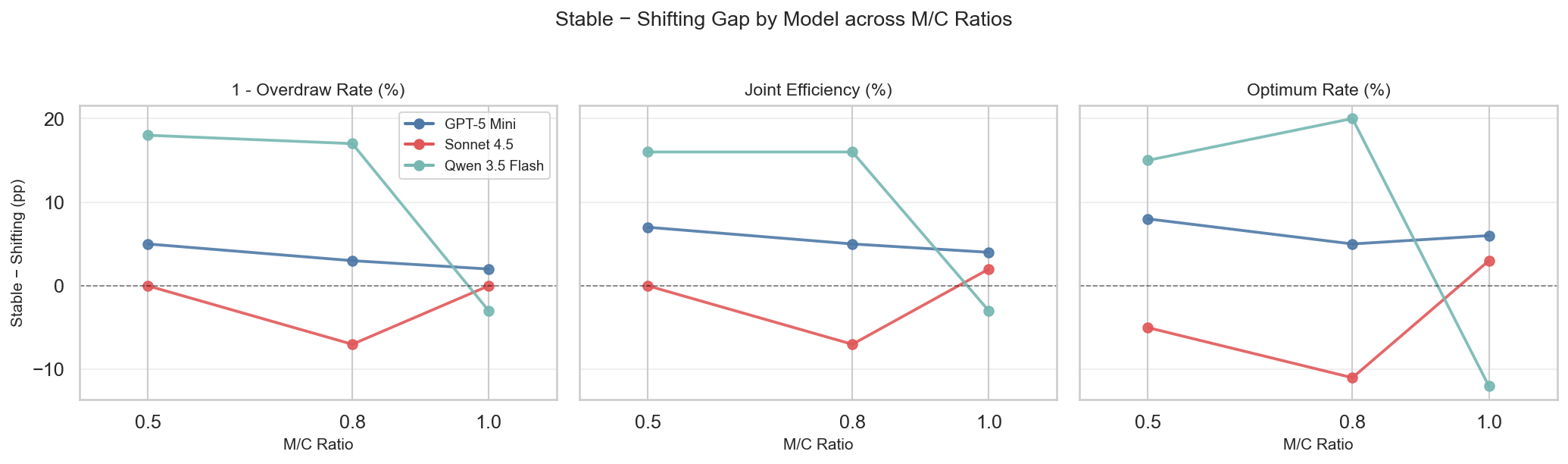}
  \caption{Stable-shifting gap by model across M/C ratios. Positive values indicate stable outperforms shifting. GPT-5 Mini and Qwen~3.5 Flash benefit from shared history, but not Sonnet~4.5.}
  \label{fig:stable_shifting}
\end{figure}

The \emph{shifting} condition, where one agent's context resets each round, degrades coordination for most models (Figure~\ref{fig:stable_shifting}). The proportion of rounds achieving the joint optimum drops by 7.7 percentage points for Qwen~3.5 Flash and 6.3 points for GPT-5 Mini (averaged across M/C ratios; Table~\ref{tab:notalk}, Appendix~\ref{app:results}), consistent with accumulated history enabling grounding repair. Yet anchoring analysis reveals that accumulated context can also become a liability. Qwen~3.5 is the clearest anomaly: at $M/C{=}1.0$, shifting substantially outperforms stable, reversing the expected pattern (Figure~\ref{fig:stable_shifting}). Among suboptimal non-overdrawn Qwen rounds at $M/C{=}1.0$ where improvement was possible, stable dyads repeat their prior allocation in 60.8\% of cases (31/51) compared to 20.8\% in shifting (5/24). Though the shifting sample is small, the direction is consistent with the hypothesis that context resets break anchoring traps. Under fully compatible goals, Qwen locks into early suboptimal splits and fails to renegotiate, while forced re-negotiation in shifting inadvertently enables better outcomes.

\begin{figure}[ht]
  \centering
  \includegraphics[width=0.8\textwidth]{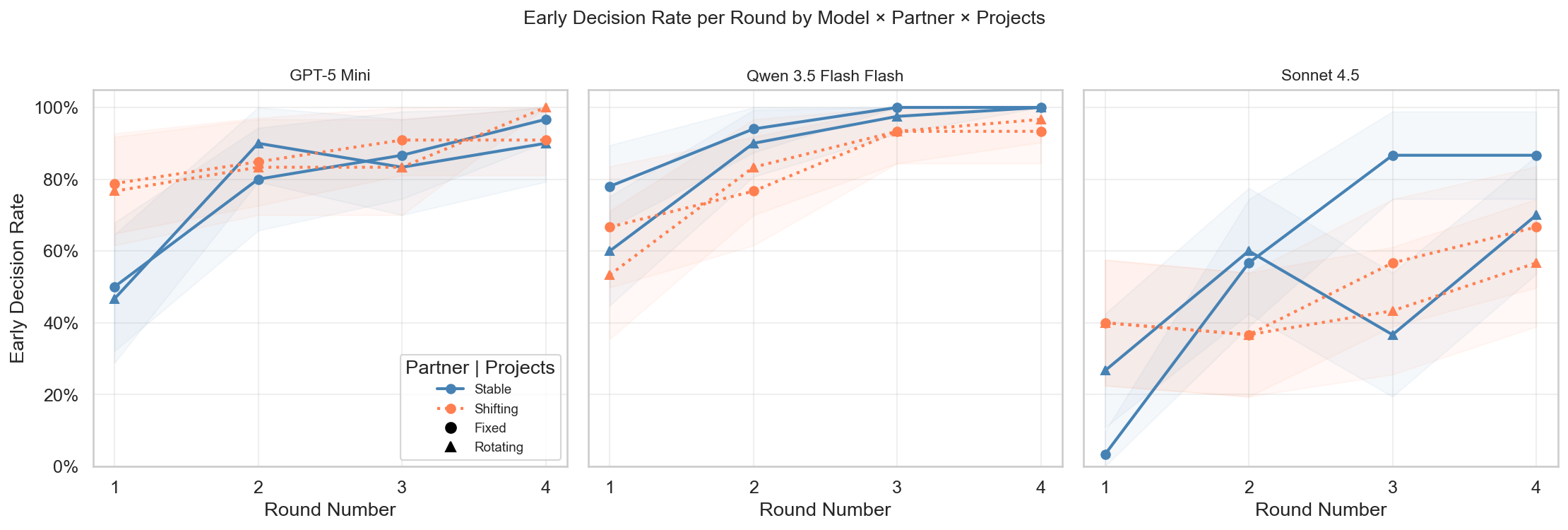}
  \caption{Rate at which an early decision is reached prior to the 5-turn conversation limit across rounds by model type, partner, and project conditions.}
  \label{fig:early_decision}
\end{figure}

Sonnet~4.5 is the broader exception to the history-as-helpful pattern, achieving slightly \emph{higher} optimum rates in the shifting condition. Differences in sustained cheap talk help explain why partner resets are less harmful for Sonnet dyads (Figure~\ref{fig:early_decision}). GPT-5 Mini and Qwen~3.5 Flash exhibit high early decision rates, submitting a resource allocation before the 5-turn conversation limit is reached. In contrast, Sonnet~4.5 has a significantly lower early decision rate and more often exhausts the available conversational turns. This prolonged engagement facilitates dynamic grounding, allowing Sonnet dyads to maintain or even improve performance with new partners each round.

\subsection{Strategy taxonomy}
\label{sec:results:strategy}

\begin{table}[ht]
\centering
\caption{Game-strategy rates across repeated rounds (non-rotating games only). Payoff alternation = higher-reward agent flips between consecutive rounds; WSLS = allocation repeat after non-overdrawn round (win-stay) / change after non-optimal round (lose-shift).}
\label{tab:strategy-main}
\small
\begin{tabular}{cccc}
\toprule
\textbf{Payoff alternation (2-rd)} & \textbf{Payoff alternation (4-rd)} & \textbf{Win-stay} & \textbf{Lose-shift} \\
\midrule
16.3\% & 2.2\% & 51.5\% & 84.3\% \\
\bottomrule
\end{tabular}
\end{table}

We classify repeated-game strategies using deterministic extraction over round outcomes and allocations (Appendix~\ref{app:strategy-taxonomy}). Table~\ref{tab:strategy-main} focuses on dyadic game-level dynamics rather than stylistic speech patterns, which are analyzed with the calibrated LLM judge in \S\ref{sec:llm-judge}. Payoff alternation, where the higher-reward agent flips between consecutive rounds, is rare (16.3\% of adjacent round pairs), and sustained 4-round alternation is almost nonexistent (2.2\%), despite being a natural coordination strategy in non-compatible scenarios where the joint optimum requires asymmetric concessions. Agents exhibit strong \emph{lose-shift} (84.3\%) and moderate \emph{win-stay} (51.5\%) behavior: they usually change allocation after overdrawn or suboptimal rounds, but they repeat prior allocations after merely non-overdrawn rounds about half the time, even when better joint strategies remain available.

\begin{figure}[ht]
  \centering
  \includegraphics[width=0.6\linewidth]{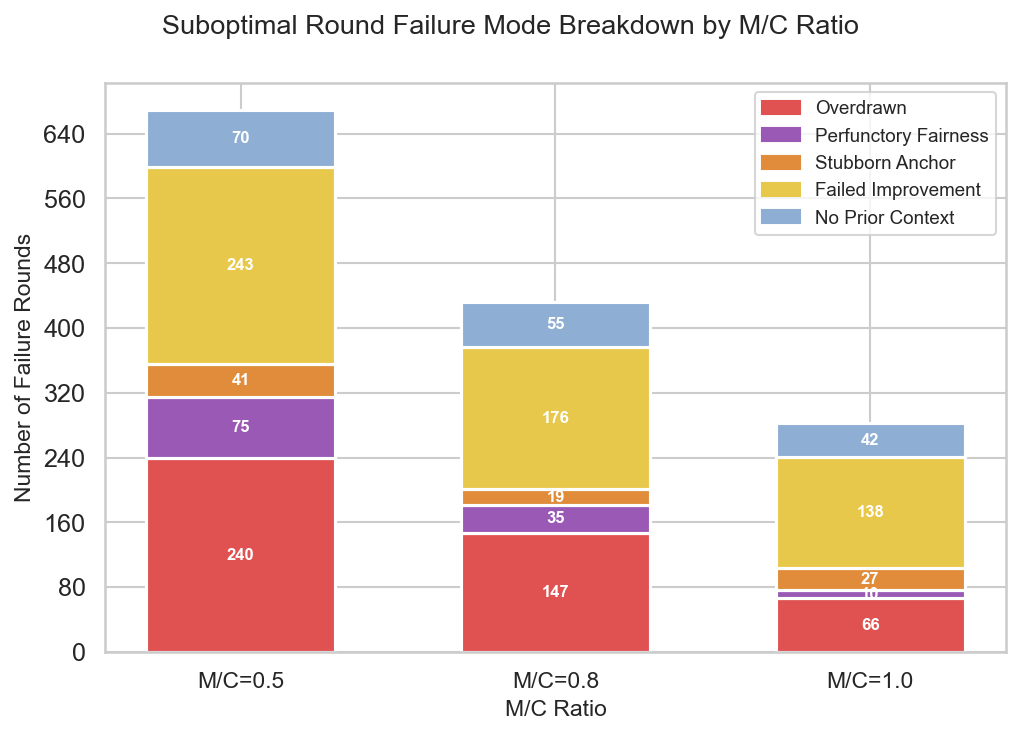}
  \caption{Failure mode breakdown for suboptimal rounds by compatibility ratio. \emph{No Prior Context} captures round-1 suboptimality before any shared history is established; \emph{Failed Improvement} captures rounds 2+ where the allocation changed but remained suboptimal. These two buckets together account for the majority of failures across all conditions. Calibrated LLM-assisted annotations further decompose these buckets (\S\ref{sec:llm-judge}, Appendix~\ref{app:judge}).}
  \label{fig:failure_modes}
\end{figure}

\subsection{Perfunctory fairness}
\label{sec:results:fairness}

We detect \emph{equal splits}---rounds where both agents request identical quantities of a shared resource---in 12.5\% of rounds. Of these, 48.8\% co-occur with suboptimal outcomes, yielding an overall \emph{perfunctory fairness} rate of 6.1\%. This pattern reflects a coordination shortcut: agents default to an equal division that is safe but ignores the asymmetric project structure that makes unequal allocations jointly better. Models differ in their reliance on this heuristic: Sonnet~4.5 produces equal splits most frequently (15.9\%, perfunctory fairness rate 7.3\%), Qwen~3.5 Flash at an intermediate rate (12.7\%, 6.7\%), and GPT-5 Mini least (10.7\%, 4.7\%). The pattern suggests agents are satisficing on surface-level fairness rather than engaging in the deeper information exchange needed to discover what each agent actually requires. Rather than asking which resources are flexible or what projects are being pursued, agents default to equal division as a socially acceptable shortcut that avoids the conversational work of grounding asymmetric needs.

\subsection{Anchoring and proposal deference}
\label{sec:results:anchoring}

\paragraph{Stubborn anchoring} In games with the same scenario, agents repeat their exact joint allocation in 46.5\% of rounds. The repeat rate is 55.1\% following non-overdrawn rounds and 0.6\% after overdraw, indicating that agents recognize and correct clear mistakes. However, repeating an allocation is not always optimal: of the 339 rounds that were suboptimal and non-overdrawn (i.e.\ failures the dyad \emph{could} have corrected), 29.2\% repeat the same allocation unchanged. This stubborn anchoring is more prevalent in stable games (40.7\%) than shifting games (15.9\%), where context resets discourage direct reuse, and increases with collaborative pressure: $M/C{=}1.0$ games exhibit a 46.0\% stubborn-anchor rate versus 28.3\% at $M/C{=}0.5$, suggesting that in compatible situations, agents have less incentive to improve upon a prior suboptimal allocation.

\paragraph{First-proposal deference} Explicit other-directed proposals (``you take $N$ of resource~$X$'') appear in 12.9\% of rounds, typically issued at the very first cheap-talk turn (median turn~1) before any information has been exchanged. In round~1 of non-rotating games, before either agent has observed an allocation outcome, 20.4\% of these proposals are followed exactly. This uninformed compliance is costly: accepted proposals lead to suboptimal outcomes almost as often as optimal ones (8.2\% vs.\ 12.2\%). Across all rounds, deference is more prevalent when scenarios change: in non-rotating games, 15.5\% of exact-match deferences produce suboptimal outcomes versus 16.9\% optimal, and the imbalance worsens in rotating games (20.9\% suboptimal vs.\ 14.4\% optimal) where per-round scenario changes make early proposals especially uninformed. Models differ in compliance: Sonnet~4.5 accepts 44\% of proposals, Qwen~3.5 Flash 32\%, and GPT-5 Mini 21\%. We read this as over-reliance on communicative satisficing rather than an outright grounding failure: consistent with Clark's principle of least collaborative effort \citep{clark1996using}, accepting a ``good enough'' first proposal is a familiar human heuristic that conserves conversational effort and avoids social friction. In our task, however, the strategy is miscalibrated, as agents accept proposals before enough information has been exchanged to assess mutual benefit, and the shortcut locks dyads into suboptimal outcomes more often than not.

\subsection{Referential binding failures}
\label{sec:results:referential}

Of the 15.7\% of rounds that result in overdraw, many follow apparently successful cheap-talk coordination: agents verbally settle on resource quantities, then submit allocations that violate the shared supply constraint. The LLM judge labels \emph{Agreement Abandonment} in 41.7\% of overdrawn rounds (Table~\ref{tab:judge-prevalence-detailed}), and the enrichment analysis in Figure~\ref{fig:judge-deltas} shows that this label is disproportionately associated with overdraw relative to non-overdrawn rounds. Manual inspection reveals two recurring subtypes (see Appendix~\ref{app:game_traces} for annotated transcripts). In \textbf{proposer amnesia}, the agent that originated a split fails to honor it: its thinking trace re-derives a new allocation from scratch with no reference to its own prior proposal. In \textbf{self-commitment abandonment}, an agent proposes a plan, the opponent accepts, and the proposer reverts to an individual-maximizing allocation once a higher-reward option becomes salient, resembling the \emph{stylistic sycophancy} of \citet{eisenstein2026mtpingeval}, where agreement language functions as social lubrication without updating the action plan. Agreements established in cheap talk appear not to be reliably retrieved as action constraints at decision time, resulting in a loss of common ground. We posit this failure results from the model failing to bind resource symbols to their earlier commitments.

Referential binding is the cognitive process of mapping a linguistic expression to a specific discourse referent. In the context of joint action \citep{clark1996using}, this mapping serves as a coordination device: once an agent confirms ``I'll take stone$\times$6,'' that quantity is anchored as a pragmatic constraint on its subsequent action. Mechanistic interpretability work has studied how LMs retrieve bound entities in context \citep{gurarieh2025mixing}, but only with simple single-query settings (``Pete loves jam, Ann loves pie. Who loves pie?''). Negotiation demands binding multiple entities simultaneously: strategies to resource-quantity pairs, quantities to agents, and prior commitments to action constraints, across a multi-turn dialogue interleaved with persuasion and counter-proposals. The failures we observe suggest that agreements established in cheap talk were not retrieved as constraints for the downstream action, resulting in the loss of common ground. Extending mechanistic binding analyses to such multi-entity, multi-turn settings is a promising direction for future research.

\subsection{Full-transparency intervention}
\label{sec:results:transparency}

To isolate how much of the coordination gap stems from information asymmetry, we ran a full-transparency intervention in which both agents' complete project specifications are revealed in the system prompt before cheap talk begins, eliminating information asymmetry entirely. We selected the Qwen~3.5~Flash $\times$ GPT-5~Mini pair for this intervention because it is the weakest-performing cross-play pair at $M/C{=}0.5$ (74.0\% efficiency, Table~\ref{tab:crossplay}), making it a conservative test of whether information asymmetry is the primary bottleneck. We ran $N{=}120$ games across all three compatibility ratios, both partner-stability conditions, and both project-rotation conditions.

Full transparency consistently reduced overdraw (12.3\%~$\rightarrow$~8.1\%, $\Delta{=}{-}4.2$~pp) and raised joint efficiency (79.6\%~$\rightarrow$~85.3\%, $\Delta{=}{+}5.7$~pp), with the sharpest gains in the stable competitive condition ($M/C{=}0.5$: overdraw $11.2\%~\rightarrow~3.8\%$, efficiency $78.1\%~\rightarrow~89.0\%$). However, the optimum rate showed no reliable improvement overall (52.9\%~$\rightarrow$~50.0\%) and even declined at $M/C{=}1.0$ (69.4\%~$\rightarrow$~64.4\%), where project requirements are already non-conflicting.

Our static process metrics distinguish which failure modes are ameliorated by shared information from those rooted in the interactive process itself. Perfunctory fairness is nearly eliminated (5.0\%~$\rightarrow$~1.7\%): when both agents can see the asymmetric project structure directly, the equal-split shortcut is visibly costly, removing the incentive for the effort-minimizing heuristic. Stubborn anchoring, by contrast, more than doubles (23.1\%~$\rightarrow$~46.7\%), suggesting the failure is driven by interaction dynamics rather than uncertainty about the partner's needs. With full information established upfront, agents converge on an early plan and then do not renegotiate even when the allocation is suboptimal. A smaller joint-transparency variant ($N{=}12$) that additionally exposed each agent's reward function achieved the highest optimum rate observed (64.6\%), suggesting that knowing the opponent's \emph{incentives}, not just resource needs, matters for closing the residual gap.

These results support the view that the coordination bottleneck is not reducible to information asymmetry. In the matched Qwen~3.5~Flash $\times$ GPT-5~Mini comparison, full transparency sharply reduces judge-labeled \emph{Misaligned Mental Models} (37.5\%~$\rightarrow$~1.0\%), even though agents still voluntarily disclose project information in over half of rounds (61.3\%). This suggests that placing partner project structure in the system prompt may be treated as a stronger shared-state prior than the same information exchanged through cheap talk. By contrast, \emph{Agreement Abandonment} is nearly unchanged (14.8\%~$\rightarrow$~13.8\%), reinforcing that commitment maintenance and speech--action consistency require interventions beyond simply exposing the hidden project state. Full per-condition outcome metrics are in Table~\ref{tab:transparency} (Appendix~\ref{app:results}).

\subsection{LLM-assisted behavioral analysis}
\label{sec:llm-judge}

\begin{figure*}[ht]
  \centering
  \begin{minipage}[t]{0.48\textwidth}
    \centering
    \includegraphics[width=\linewidth]{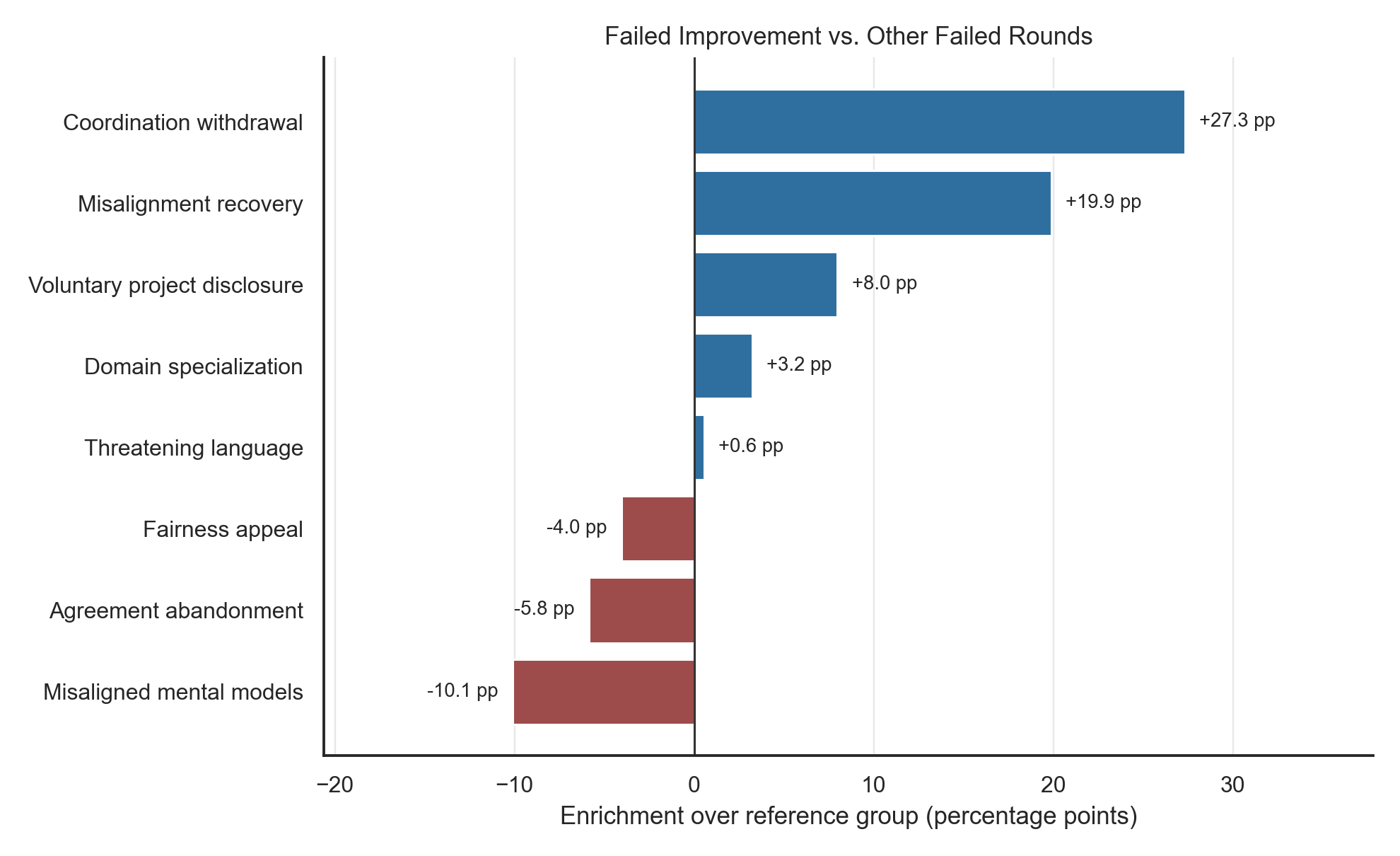}
    \smallskip
    \small (a) Failed improvement vs.\ other suboptimal rounds
  \end{minipage}
  \hfill
  \begin{minipage}[t]{0.48\textwidth}
    \centering
    \includegraphics[width=\linewidth]{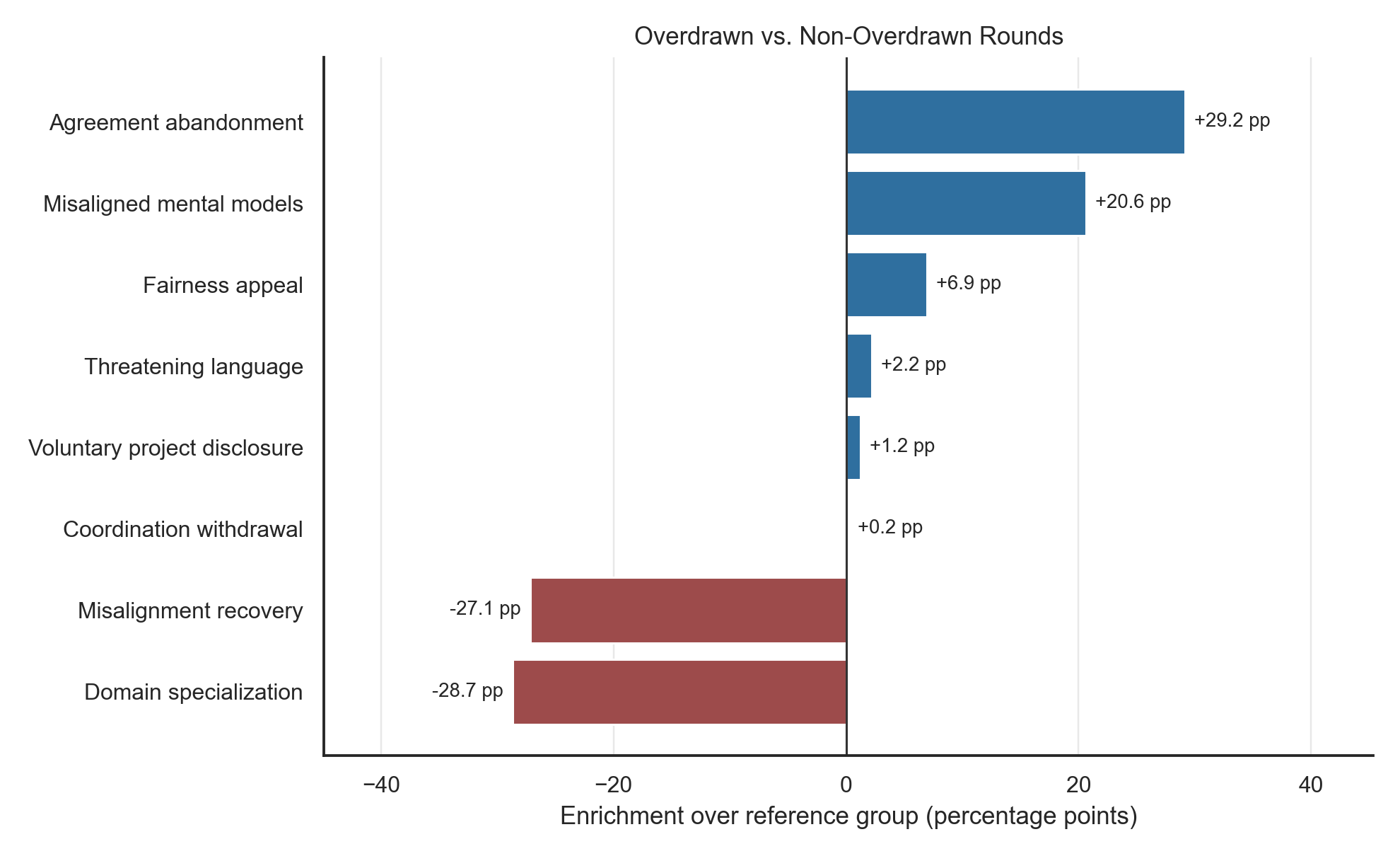}
    \smallskip
    \small (b) Overdrawn vs.\ non-overdrawn rounds
  \end{minipage}
  \caption{Judge-label enrichment in actionable failure regions. Bars show percentage-point differences in calibrated judge-label prevalence between the focal condition and its comparison set. Positive values indicate labels disproportionately associated with the focal failure condition, suggesting candidate targets for future intervention rather than causal effects.}
  \label{fig:judge-deltas}
\end{figure*}

Our main quantitative results are based on objective game-level metrics, including overdraw rate, allocation efficiency, optimum rate, and failed improvement across repeated interaction. The LLM-judge pipeline is used to explain these metrics rather than define them: it provides a secondary, correlational lens on the communication patterns associated with overdrawn, suboptimal, or non-improving rounds.

\begin{table}[ht]
\centering
\caption{LLM judge label prevalence across the 720-game cohort ($N=2{,}880$ rounds). Labels are not mutually exclusive.}
\label{tab:judge-main-prevalence}
\scriptsize
\setlength{\tabcolsep}{4pt}
\begin{tabular}{lrr}
\toprule
\textbf{Label} & \textbf{Rounds} & \textbf{\%} \\
\midrule
Misaligned mental models & 616 & 21.4 \\
Agreement abandonment & 492 & 17.1 \\
Coordination withdrawal & 940 & 32.6 \\
Domain specialization & 742 & 25.8 \\
Misalignment recovery & 685 & 23.8 \\
Voluntary project disclosure & 1605 & 55.7 \\
Fairness appeal & 502 & 17.4 \\
Threatening language & 68 & 2.4 \\
\bottomrule
\end{tabular}
\vskip -0.1in
\end{table}

We ran this pipeline over all 720 game traces (details in Appendix~\ref{app:judge}). Table~\ref{tab:judge-main-prevalence} summarizes the prevalence of five core round-level labels and three auxiliary speech tags. To calibrate the automated annotations, we compared the labels from our judge pipeline against human annotations on a 40-round subset spanning 10 games. Agreement is at or above Cohen's Kappa $\kappa{\approx}0.40$ for every label except \emph{Misalignment Recovery} ($\kappa=0.30$; full calibration in Table~\ref{tab:judge-human-kappa}). These agreement levels support using the judge outputs as calibrated, exploratory process measurements for interpreting objective outcome metrics and the label-prevalence shifts in Figure~\ref{fig:judge-deltas}, rather than as ground-truth behavioral labels.

While the rule-based failure breakdown in Figure~\ref{fig:failure_modes} shows where suboptimal outcomes sit structurally, it does not by itself reveal which conversational mechanisms characterize each bucket. We therefore compare calibrated judge-label prevalence inside two actionable failure regions against targeted comparison sets (Figure~\ref{fig:judge-deltas}). Compared with other suboptimal rounds, failed-improvement rounds show the largest increase in \emph{Coordination Withdrawal}, suggesting that agents often avoid or prematurely narrow negotiation even when continued exchange could repair the trajectory. By contrast, overdrawn rounds show higher rates of \emph{Misaligned Mental Models} and \emph{Agreement Abandonment}, consistent with agents either losing track of the shared state or failing to align final actions with prior public commitments.

\section{Discussion}
\label{sec:discussion}

Together, our three baselines decompose the coordination gap and localize the bottleneck. The oracle baseline rules out individual reasoning limitations: agents can identify optimal allocations in isolation, so the gap is not attributable to bounded rationality. The no-talk baseline establishes that communication is necessary: without cheap talk, joint efficiency roughly triples under competition (\S\ref{sec:results:notalk}), confirming that information exchange is a prerequisite for coordination. The full-transparency intervention (\S\ref{sec:results:transparency}) then isolates what shared information can and cannot fix. Transparency reduces overdraw and dissolves perfunctory fairness, confirming those are partly attributable to uncertainty about the partner's portfolio. The optimum rate, however, does not reliably improve, and stubborn anchoring intensifies. The failures that persist under full information, namely anchoring, commitment-action decoupling, and referential binding, are therefore not reducible to information asymmetry. They are grounded in the interactive process itself: joint plan formation, commitment maintenance, and execution under cheap talk. The grounding gap is not a gap in knowledge or reasoning capacity; it is a gap in the ability to coordinate action across turns.

\paragraph{Connection to grounding theory} Our failure modes map onto established mechanisms. Anchoring reflects a breakdown of Clark and Brennan's contribution--acceptance cycle \citep{clark1991grounding}: proposals enter common ground without verification, and in Traum's terms \citep{traum1994computational}, agents over-initiate while under-acknowledging. First offers pull final outcomes even in LLM negotiations \citep{tversky1974judgment, takenami2025anchoring}, and this resembles sycophancy \citep{perez2023sycophancy, sharma2024sycophancy}; while \citet{eisenstein2026mtpingeval} found agents can reject incorrect proposals in their chess task, our agents fail to establish the mutual information sufficient for the task at hand. Perfunctory fairness reflects minimizing \emph{individual} rather than \emph{joint} effort \citep{clark1986referring}, and the win-stay/lose-shift pattern mirrors the Pavlov strategy \citep{nowak1993strategy}, effective reactively but unable to improve on suboptimal successes.

\paragraph{Limitations \& Future Work} Our study is limited to dyadic, text-only interactions between three models; flagship models and human-human dyads remain untested. While our project structure adds complexity over scalar payoff matrices, it remains a stylized abstraction. Specifically, our environment assumes symmetric power dynamics and enforces a fixed 5-turn communication window, whereas real-world negotiation frequently involves hierarchical leverage and open-ended dialogue. Our process metrics use a LLM judge pipeline, which can be unreliable and better calibrated with more human annotators. Further isolating grounding breakdowns from rational defection, and expanding this framework to multi-party coalitions ($N>2$) and strategic perturbations (e.g., agents with adversarial personas), remain critical frontiers for future work.

Our failure mode analysis surfaces targeted directions for future intervention work. The failed-improvement results point to mechanisms that keep attempted negotiations active rather than avoidant: agents need scaffolds that encourage continued clarification, counter-proposal, and repair instead of settling for a heuristic fallback. The overdraw results point to two complementary targets: better theory-of-mind state tracking to reduce misaligned mental models, and commitment scaffolds that reduce the speech--action gap before final allocation. Perfunctory fairness, meanwhile, is less about equity than about a low-effort heuristic that short-circuits the search for better complementary splits; richer structured information exchange, as demonstrated by our transparency intervention, largely dissolves this pattern. Finally, our shifting-vs.-stable comparison reveals that a consistent partner is helpful for coordination: efficiency is systematically higher in stable dyads, suggesting that agent architectures equipped with persistent partner models or episodic interaction summaries may reduce grounding failures in multi-session deployments.

\section{Conclusion}
\label{sec:conclusion}

We introduced an iterated negotiation game that decomposes multi-agent coordination failure into measurable components. The oracle baseline shows that the gap is not attributable to individual reasoning limitations; the no-talk baseline shows that communication is necessary; and the full-transparency intervention shows that information exchange alone is insufficient. Together, these results turn dynamic grounding from a human-inspired account of communication into a falsifiable mechanism for agent failure and improvement. Even when agents reason individually, communicate freely, and see each other's private goals, the residual gap lies in maintaining common ground through joint plan formation, commitment maintenance, and execution across turns. This decomposition identifies intervention targets: structured information exchange reduces uncertainty-driven shortcuts such as perfunctory fairness, but future systems still need explicit grounding protocols, commitment scaffolds, and persistent partner models. With LLM deployment shifting toward autonomous agents, effective communication remains essential but underdeveloped. Our controlled paradigm isolates the multi-agent coordination gap from failures of individual reasoning or information access, showing that agents must update shared plans, honor commitments, and carry agreements into action.

\section*{Impact Statement}
As AI systems become personal delegates, a central trust question is whether they can represent users faithfully in interactions with other agents. This requires more than solving isolated tasks: assistants must maintain user preferences, negotiate under uncertainty, honor commitments, and update shared plans without losing sight of the person they represent. These capabilities are dual-use: stronger agent-agent coordination could make harmful or manipulative collectives more effective, make inter-agent agreements harder for humans to oversee or interrupt, and increase privacy leakage if agents disclose user goals or constraints too freely. Our controlled synthetic setting exposes failures that arise when agents turn private goals into joint action through conversation. Surfacing these vulnerabilities now can inform verifiable guardrails before such delegates are deployed in open-ended applications.

\section*{LLM Use Statement}
Claude Code (Anthropic) was used as a coding assistant for game engine development, data processing scripts, and the generation of appendix sections (scenario listings, prompt templates, and LLM-judge examples). Perplexity, Gemini, and Claude were used for synthesis during literature review. All outputs were reviewed and edited by the authors. The experimental results, analysis, and scientific conclusions are the authors' own.

\bibliography{arxiv_submission}
\bibliographystyle{arxiv_submission}

\appendix
\setlength{\parindent}{0pt}
\setlength{\parskip}{0.5ex}

\section{Interactivity level analysis}
\label{app:interactivity}

\citet{eisenstein2026mtpingeval} propose a formal hierarchy of \emph{interactivity levels} for collaborative games with private information, where level~$k$ indicates that $k$ messages must be exchanged before a player can produce a correct answer. Their framework assumes a shared objective; we use it to establish a \emph{lower bound} on the communication required in our game, where agents are instructed to maximize their own individual reward.

Let $X_1$ and $X_2$ denote each agent's private project information. Both agents must submit allocation decisions. We claim our game is at least level-3 interactive assuming the players are initially unaware of the $M/C$ ratio (degree of goal compatibility):

\begin{enumerate}[leftmargin=*, itemsep=2pt]
  \item Agent~2 shares their project requirements: $f_2(X_2)$.
  \item Agent~1, now informed of both sets of projects, computes and proposes a specific compatible allocation: $f_1(X_1, S_2)$.
  \item Agent~2 verifies the proposal against their own projects and confirms or counter-proposes: $f_2(X_2, S_1)$.
\end{enumerate}

This third turn is necessary regardless of the compatibility ratio. Even when $M/C = 1.0$, multiple allocation plans may satisfy both agents' individual optima, and not all combinations are supply-compatible---without explicit coordination on \emph{which} plan to execute, overdraw can occur. When $M/C < 1.0$, the candidate plans additionally involve trade-offs between agents' rewards, making the confirmation step more contentious.

These three turns represent only the information-exchange minimum. Since agents maximize individual rewards, an additional \emph{bargaining} dimension arises that the interactivity framework does not capture: agents must negotiate over who concedes what, a process whose difficulty scales with goal incompatibility. The information-exchange minimum (level~3) is constant across compatibility conditions; the compatibility ratio modulates the difficulty of the bargaining problem that follows.

\section{LLM model configuration details}
\label{app:models}

Table~\ref{tab:model-config} summarizes the API configuration for the three main-cohort models and additional exploratory models used in follow-up runs. All models use streaming responses. A 1-second cooldown is enforced between consecutive API calls, with up to 10 retries using exponential backoff (base 2s, max 120s) for rate-limit errors. Malformed JSON responses are retried up to 3 times per decision.

\paragraph{Agent errors} Across 950 games (3,800 rounds), 29.4\% of games contain at least one logged error event (496 total; 0.52 per game). Importantly, \texttt{validation\_error} and \texttt{output\_format\_warning} events (436 combined, 87.9\%) are \emph{retriable}: the engine re-prompts the agent up to 3 times before escalating. A \texttt{heuristic\_fallback} entry is the definitive signal that an agent failed to produce a valid allocation after all retries---only 20 such events occur in the dataset, all from Qwen~3.6 Plus (an exploratory model outside the main reported set), attributable to sustained rate-limit failures exhausting the retry budget. \texttt{decision\_auto\_filled} (20 events, also Qwen~3.6 Plus) indicates rounds where a default zero-allocation was substituted as a result. Rounds containing heuristic fallbacks are excluded from all analyses. All three main models (Sonnet~4.5, GPT-5 Mini, Qwen~3.5 Flash) recovered within the retry budget on every round; their per-round error rates reflect only retriable validation errors: Qwen~3.5 Flash 0.073, Sonnet~4.5 0.071, GPT-5 Mini 0.047. The exploratory models differ in error profile: MiniMax M2.7 has a higher retriable validation-error rate (0.090), while Qwen~3.6 Plus has the only sustained rate-limit fallbacks despite a lower aggregate per-round error rate (0.048).

\begin{table}[h]
\centering
\small
\caption{Model configurations used in experiments. All models accessed via their respective provider APIs.}
\label{tab:model-config}
\begin{tabular}{lllcc}
\toprule
\textbf{Model} & \textbf{Provider} & \textbf{Model string} & \textbf{Temp.} & \textbf{Max tokens} \\
\midrule
Claude 4.5 Sonnet & Anthropic & \texttt{claude-sonnet-4-5-20250929} & 0.7 & 8192 \\
GPT-5 Mini & OpenAI & \texttt{gpt-5-mini-2025-08-07} & --- & 16000 \\
Qwen 3.5 Flash & OpenRouter & \texttt{qwen/qwen3.5-flash} & 0.7 & 16384 \\
\midrule
\multicolumn{5}{l}{\emph{Additional models (V5--V6 experiments):}} \\
Gemini 3 Flash & Google & \texttt{gemini-3-flash-preview} & 0.7 & 16384 \\
Haiku 4.5 & Anthropic & \texttt{claude-haiku-4-5-20251001} & 0.7 & 8192 \\
GPT-5.4 Mini & OpenAI & \texttt{gpt-5.4-mini} & --- & 16000 \\
Nemotron & OpenRouter & \texttt{nvidia/nemotron-3-nano-30b-a3b:free} & 0.7 & 4096 \\
\bottomrule
\end{tabular}
\end{table}

\noindent GPT-5 Mini and GPT-5.4 Mini use OpenAI's \texttt{max\_completion\_tokens} parameter and do not support explicit temperature settings. All other models use the standard \texttt{max\_tokens} parameter with temperature 0.7. Anthropic models support prompt caching; reasoning content (chain-of-thought) is extracted and stored separately from speech for all models that produce it.

\section{Generated scenarios}
\label{app:scenarios}
\subsection{Scenario Themes}
To ensure model generalizability and prevent overfitting to specific resource names, we employ thematic randomization. Each game round is assigned a random theme that renames the abstract resources (r1, r2, r3). Table~\ref{tab:themes} lists the available themes and their corresponding resource names.
\begin{table}[H]
\centering
\small
\begin{tabular}{llll}
\toprule
Theme & r1 (Cost 1.0) & r2 (Cost 1.5) & r3 (Cost 3.0) \\
\midrule
Medieval & wood & stone & gold \\
Space & titanium & crystal & plasma \\
Undersea & coral & pearl & trident \\
Steampunk & copper & brass & aether \\
Jungle & bamboo & vine & amber \\
Desert & sandstone & glass & ruby \\
Arctic & ice & fur & diamond \\
Volcanic & obsidian & basalt & magma \\
Cyberpunk & silicon & fiber & quantum \\
Fairy\_tale & pixie\_dust & moonstone & starlight \\
\bottomrule
\end{tabular}
\caption{Thematic resource mappings.}
\label{tab:themes}
\end{table}
\subsection{Scenario Pool Details}
All scenarios in the pool share a common resource environment: Supply (r1: 10, r2: 10, r3: 6), Costs (r1: 1.0, r2: 1.5, r3: 3.0), and individual Agent Budget (18.0).
\subsubsection{Compatibility Ratio $M/C \approx 0.5$}
\paragraph{Scenario gen\_012 ($M/C = 0.54$)}
\begin{table}[H]
\centering
\small
\begin{tabular}{lp{4cm}r | lp{4cm}r}
\toprule
\multicolumn{3}{c}{\textbf{Agent 1 Projects}} & \multicolumn{3}{c}{\textbf{Agent 2 Projects}} \\
Name & Requirements & Reward & Name & Requirements & Reward \\
\midrule
project\_a & r2x3 & 9 & project\_a & r3x3 & 1 \\
project\_b & r3x2, r2x3 & 4 & project\_b & r1x1, r2x3 & 9 \\
project\_c & r3x2, r1x1 & 1 & project\_c & r2x2 & 4 \\
\bottomrule
\end{tabular}
\end{table}

\paragraph{Scenario gen\_053 ($M/C = 0.50$)}
\begin{table}[H]
\centering
\small
\begin{tabular}{lp{4cm}r | lp{4cm}r}
\toprule
\multicolumn{3}{c}{\textbf{Agent 1 Projects}} & \multicolumn{3}{c}{\textbf{Agent 2 Projects}} \\
Name & Requirements & Reward & Name & Requirements & Reward \\
\midrule
project\_a & r3x3, r2x5 & 3 & project\_a & r1x3, r3x4 & 2 \\
project\_b & r1x4, r3x3 & 3 & project\_b & r1x2, r2x5 & 3 \\
project\_c & r1x5 & 6 & project\_c & r1x3, r3x2 & 6 \\
\bottomrule
\end{tabular}
\end{table}

\paragraph{Scenario gen\_062 ($M/C = 0.53$)}
\begin{table}[H]
\centering
\small
\begin{tabular}{lp{4cm}r | lp{4cm}r}
\toprule
\multicolumn{3}{c}{\textbf{Agent 1 Projects}} & \multicolumn{3}{c}{\textbf{Agent 2 Projects}} \\
Name & Requirements & Reward & Name & Requirements & Reward \\
\midrule
project\_a & r2x6 & 3 & project\_a & r3x1, r1x4 & 7 \\
project\_b & r3x5 & 1 & project\_b & r2x3, r1x6 & 4 \\
project\_c & r2x1, r1x5 & 9 & project\_c & r2x5, r1x1 & 9 \\
\bottomrule
\end{tabular}
\end{table}

\paragraph{Scenario gen\_104 ($M/C = 0.53$)}
\begin{table}[H]
\centering
\small
\begin{tabular}{lp{4cm}r | lp{4cm}r}
\toprule
\multicolumn{3}{c}{\textbf{Agent 1 Projects}} & \multicolumn{3}{c}{\textbf{Agent 2 Projects}} \\
Name & Requirements & Reward & Name & Requirements & Reward \\
\midrule
project\_a & r1x2 & 4 & project\_a & r2x4, r3x3 & 1 \\
project\_b & r3x3, r1x3 & 7 & project\_b & r1x1 & 2 \\
project\_c & r1x6, r2x6 & 2 & project\_c & r3x4, r1x2 & 5 \\
\bottomrule
\end{tabular}
\end{table}

\subsubsection{Compatibility Ratio $M/C \approx 0.8$}
\paragraph{Scenario gen\_001 ($M/C = 0.83$)}
\begin{table}[H]
\centering
\small
\begin{tabular}{lp{4cm}r | lp{4cm}r}
\toprule
\multicolumn{3}{c}{\textbf{Agent 1 Projects}} & \multicolumn{3}{c}{\textbf{Agent 2 Projects}} \\
Name & Requirements & Reward & Name & Requirements & Reward \\
\midrule
project\_a & r3x2, r2x6 & 8 & project\_a & r2x2, r3x2 & 5 \\
project\_b & r3x5, r1x2 & 5 & project\_b & r3x4 & 5 \\
project\_c & r3x1, r2x2 & 5 & project\_c & r1x6 & 10 \\
\bottomrule
\end{tabular}
\end{table}

\paragraph{Scenario gen\_006 ($M/C = 0.80$)}
\begin{table}[H]
\centering
\small
\begin{tabular}{lp{4cm}r | lp{4cm}r}
\toprule
\multicolumn{3}{c}{\textbf{Agent 1 Projects}} & \multicolumn{3}{c}{\textbf{Agent 2 Projects}} \\
Name & Requirements & Reward & Name & Requirements & Reward \\
\midrule
project\_a & r3x1, r2x2 & 6 & project\_a & r2x2, r1x1 & 6 \\
project\_b & r3x5, r1x3 & 5 & project\_b & r1x3 & 6 \\
project\_c & r2x2 & 6 & project\_c & r3x5, r1x1 & 1 \\
\bottomrule
\end{tabular}
\end{table}

\paragraph{Scenario gen\_017 ($M/C = 0.83$)}
\begin{table}[H]
\centering
\small
\begin{tabular}{lp{4cm}r | lp{4cm}r}
\toprule
\multicolumn{3}{c}{\textbf{Agent 1 Projects}} & \multicolumn{3}{c}{\textbf{Agent 2 Projects}} \\
Name & Requirements & Reward & Name & Requirements & Reward \\
\midrule
project\_a & r2x1, r3x3 & 2 & project\_a & r2x3 & 3 \\
project\_b & r1x1, r2x3 & 3 & project\_b & r2x1, r3x4 & 2 \\
project\_c & r3x5 & 6 & project\_c & r3x3, r1x4 & 6 \\
\bottomrule
\end{tabular}
\end{table}

\paragraph{Scenario gen\_021 ($M/C = 0.83$)}
\begin{table}[H]
\centering
\small
\begin{tabular}{lp{4cm}r | lp{4cm}r}
\toprule
\multicolumn{3}{c}{\textbf{Agent 1 Projects}} & \multicolumn{3}{c}{\textbf{Agent 2 Projects}} \\
Name & Requirements & Reward & Name & Requirements & Reward \\
\midrule
project\_a & r1x2 & 1 & project\_a & r3x3, r1x3 & 6 \\
project\_b & r2x1, r1x3 & 6 & project\_b & r2x3 & 6 \\
project\_c & r3x3 & 6 & project\_c & r1x2, r2x4 & 5 \\
\bottomrule
\end{tabular}
\end{table}

\paragraph{Scenario gen\_022 ($M/C = 0.83$)}
\begin{table}[H]
\centering
\small
\begin{tabular}{lp{4cm}r | lp{4cm}r}
\toprule
\multicolumn{3}{c}{\textbf{Agent 1 Projects}} & \multicolumn{3}{c}{\textbf{Agent 2 Projects}} \\
Name & Requirements & Reward & Name & Requirements & Reward \\
\midrule
project\_a & r1x6 & 8 & project\_a & r3x2, r2x6 & 8 \\
project\_b & r3x4 & 4 & project\_b & r1x2, r3x2 & 1 \\
project\_c & r2x2, r3x1 & 4 & project\_c & r3x3 & 6 \\
\bottomrule
\end{tabular}
\end{table}

\subsubsection{Compatibility Ratio $M/C \approx 1.0$}
\paragraph{Scenario gen\_000 ($M/C = 1.00$)}
\begin{table}[H]
\centering
\small
\begin{tabular}{lp{4cm}r | lp{4cm}r}
\toprule
\multicolumn{3}{c}{\textbf{Agent 1 Projects}} & \multicolumn{3}{c}{\textbf{Agent 2 Projects}} \\
Name & Requirements & Reward & Name & Requirements & Reward \\
\midrule
project\_a & r3x3, r1x1 & 5 & project\_a & r2x4, r1x4 & 6 \\
project\_b & r3x1, r1x5 & 1 & project\_b & r2x4, r3x1 & 3 \\
project\_c & r2x5, r3x1 & 1 & project\_c & r1x3, r3x3 & 6 \\
\bottomrule
\end{tabular}
\end{table}

\paragraph{Scenario gen\_001 ($M/C = 1.00$)}
\begin{table}[H]
\centering
\small
\begin{tabular}{lp{4cm}r | lp{4cm}r}
\toprule
\multicolumn{3}{c}{\textbf{Agent 1 Projects}} & \multicolumn{3}{c}{\textbf{Agent 2 Projects}} \\
Name & Requirements & Reward & Name & Requirements & Reward \\
\midrule
project\_a & r2x5, r3x3 & 7 & project\_a & r2x2 & 4 \\
project\_b & r1x1, r3x2 & 10 & project\_b & r2x4 & 10 \\
project\_c & r3x1 & 4 & project\_c & r3x1, r1x5 & 8 \\
\bottomrule
\end{tabular}
\end{table}

\paragraph{Scenario gen\_002 ($M/C = 1.00$)}
\begin{table}[H]
\centering
\small
\begin{tabular}{lp{4cm}r | lp{4cm}r}
\toprule
\multicolumn{3}{c}{\textbf{Agent 1 Projects}} & \multicolumn{3}{c}{\textbf{Agent 2 Projects}} \\
Name & Requirements & Reward & Name & Requirements & Reward \\
\midrule
project\_a & r3x3, r2x4 & 1 & project\_a & r3x3, r2x3 & 10 \\
project\_b & r3x1, r1x6 & 10 & project\_b & r3x2, r1x3 & 5 \\
project\_c & r1x1, r3x3 & 10 & project\_c & r3x6 & 8 \\
\bottomrule
\end{tabular}
\end{table}

\paragraph{Scenario gen\_006 ($M/C = 1.00$)}
\begin{table}[H]
\centering
\small
\begin{tabular}{lp{4cm}r | lp{4cm}r}
\toprule
\multicolumn{3}{c}{\textbf{Agent 1 Projects}} & \multicolumn{3}{c}{\textbf{Agent 2 Projects}} \\
Name & Requirements & Reward & Name & Requirements & Reward \\
\midrule
project\_a & r3x2 & 4 & project\_a & r1x2 & 4 \\
project\_b & r2x1 & 2 & project\_b & r2x2, r3x1 & 2 \\
project\_c & r1x6, r2x3 & 9 & project\_c & r3x1, r2x3 & 10 \\
\bottomrule
\end{tabular}
\end{table}

\paragraph{Scenario gen\_010 ($M/C = 1.00$)}
\begin{table}[H]
\centering
\small
\begin{tabular}{lp{4cm}r | lp{4cm}r}
\toprule
\multicolumn{3}{c}{\textbf{Agent 1 Projects}} & \multicolumn{3}{c}{\textbf{Agent 2 Projects}} \\
Name & Requirements & Reward & Name & Requirements & Reward \\
\midrule
project\_a & r1x5, r2x4 & 8 & project\_a & r2x4 & 4 \\
project\_b & r3x4, r2x2 & 6 & project\_b & r3x6 & 8 \\
project\_c & r3x3, r2x2 & 8 & project\_c & r1x2, r2x3 & 2 \\
\bottomrule
\end{tabular}
\end{table}

\paragraph{Scenario gen\_014 ($M/C = 1.00$)}
\begin{table}[H]
\centering
\small
\begin{tabular}{lp{4cm}r | lp{4cm}r}
\toprule
\multicolumn{3}{c}{\textbf{Agent 1 Projects}} & \multicolumn{3}{c}{\textbf{Agent 2 Projects}} \\
Name & Requirements & Reward & Name & Requirements & Reward \\
\midrule
project\_a & r3x3, r2x1 & 5 & project\_a & r3x1, r2x2 & 10 \\
project\_b & r1x3 & 10 & project\_b & r1x3, r3x1 & 10 \\
project\_c & r3x3, r1x1 & 10 & project\_c & r3x4 & 6 \\
\bottomrule
\end{tabular}
\end{table}


\section{Detailed game prompts}
\label{app:prompts}

This appendix provides the complete prompt templates used in our negotiation game. All prompts are defined in \texttt{backend/agents/prompts.py} and dynamically instantiated based on game configuration. We present prompts for the project-based reward system (schema version~5+), which is used for all experiments reported in the main text.

\subsection{System prompt structure}
\label{app:prompts:system}

Each agent receives a system-level prompt at the start of the game that remains constant throughout all rounds. The prompt consists of four main sections: game rules, project specifications, response format, and strategic context.

\subsubsection{Game rules}

\begin{verbatim}
## Game Rules
- Resources available: {resource_types}
- Resource supply per round: {resource_supply}
- Resource costs: {resource_costs}
- Your budget per round: ${agent_budget}
- You may purchase at most {max_types} different resource types per round.
- Resources do NOT carry over between rounds — each round starts fresh with a new budget and new supply.
- If total demand for ANY resource exceeds supply, the round is ANNULLED and both parties get 0 reward.
\end{verbatim}

\textbf{Example instantiation} (stable, collaborative condition):
\begin{verbatim}
- Resources available: wood, stone, gold
- Resource supply per round: {"wood": 10, "stone": 10, "gold": 6}
- Resource costs: {"wood": 1, "stone": 1.5, "gold": 3}
- Your budget per round: $18
- You may purchase at most 2 different resource types per round.
\end{verbatim}

\subsubsection{Project specifications}

\begin{verbatim}
## Projects
You have 3 projects. Each project requires specific resources per run and yields a reward per run.
The other party has their own projects with different requirements and rewards.
Your reward = sum of (runs × reward) for each project you can complete with your purchased resources.
IMPORTANT: Unspent money has NO value — only completed project runs count towards your score.
NOTE: If you purchase resources but lack the full set required to complete at least one run of any project, your reward is 0 (this is NOT an annulment — the round proceeds normally, you simply completed no projects).
{projects_info}
{opponent_projects_info}
\end{verbatim}

\textbf{Example instantiation} (agent with complementary projects):
\begin{verbatim}
  - project_a: requires [stone×2], reward = 6/run
  - project_b: requires [wood×5, gold×3], reward = 15/run
  - project_c: requires [stone×4], reward = 10/run
\end{verbatim}

\subsubsection{Decision format}

\begin{verbatim}
## Decision Format
Submit a JSON object with your resource purchases AND project allocations:
{"wood": 3, "gold": 1, "projects": {"project_a": 1, "project_b": 2}}
The "projects" field specifies how many times to run each project.
If omitted, the engine assigns resources to projects automatically by prioritizing projects in the order presented to you.
\end{verbatim}

\subsubsection{Response format (thinking enabled)}

For agents with private thinking scratchpads (all experiments in main text):

{\small
\begin{verbatim}
## Response Format
You MUST always respond with a JSON object containing exactly three
fields:
{
  "thinking": "your private reasoning (hidden from the other party)",
  "speech": "your message to the other party (visible to them)",
  "action": null
}

- Set "action" to null while you want to keep chatting.
- Set "action" to a purchase object with project allocation to
  finalize.
- "speech" MUST NOT be empty — always say something to the other
  party.

Example — chatting:
{"thinking": "They might want gold. I should bluff.",
 "speech": "I'm thinking about gold. What about you?",
 "action": null}

Example — purchasing:
{"thinking": "I'll buy wood for my projects.",
 "speech": "Good luck!",
 "action": {"wood": 5, "stone": 2,
            "projects": {"project_a": 1, "project_b": 1}}}

IMPORTANT: Respond with ONLY the JSON object. No text before or after it.

I will guide you through each phase with instructions.
\end{verbatim}

For Anthropic models, an additional suffix is appended to enforce JSON compliance:
\begin{verbatim}
IMPORTANT: You MUST respond with ONLY a valid JSON object. No other text before or after. The JSON must have exactly three fields:
'thinking' (string), 'speech' (string), and 'action' (null or object).
\end{verbatim}
}

\subsubsection{Strategic context}

The final section varies by experimental condition:

\textbf{Stable partners} (default):
\begin{verbatim}
## Your Situation
- Your goal: maximize your cumulative reward across all rounds
- The other party is also purchasing from the same shared pool.
\end{verbatim}

\textbf{Shifting partners} (partner context resets each round):
\begin{verbatim}
## Your Situation
- Your goal: maximize your cumulative reward across all rounds
- You are playing against a DIFFERENT opponent each round.
\end{verbatim}

\textbf{Project sharing condition} (intervention):
\label{app:prompts:sharing}
\begin{verbatim}
IMPORTANT — As a FIRST step in each round's cheap talk, share your project details (names, resource requirements, rewards) with the other party and ask them to share theirs.
\end{verbatim}

\textbf{Theory-of-mind condition} (intervention):
\label{app:prompts:tom}
\begin{verbatim}
IMPORTANT — During each round's cheap talk, actively reason about the other party's goals and which projects they might be trying to run. Consider what resources they need, how their interests align or conflict with yours, and how you can use this understanding to inform your strategy.
\end{verbatim}

Additionally, in the Theory-of-mind condition, the following reminder is appended to the Turn~0 cheap talk prompt:
\begin{verbatim}
Reminder: Actively reason about the other party's goals and which projects they might be trying to run. Consider what resources they need and how their interests align or conflict with yours.
\end{verbatim}

\subsection{User prompts: cheap talk phase}
\label{app:prompts:cheaptalk}

After the system prompt, agents receive user-level prompts that guide them through each phase of the game. The cheap talk phase uses alternating prompts based on turn number.

\subsubsection{Initial cheap talk prompt (Turn 0)}

\begin{verbatim}
--- CHEAP TALK PHASE (Round {round_number}/{num_rounds}) ---
Exchange messages with the other party ({turns_info}).
Respond with JSON. Set "action" to null to keep chatting, or to a purchase object to finalize.

[If opponent spoke first:]
The other party said: "{opponent_message}"
\end{verbatim}

Where \texttt{turns\_info} is either ``up to 5 exchanges this round'' (default) or ``as much as required'' if \texttt{cheap\_talk\_turns=0}.

\textbf{Example for Round 1 of 4-round game:}

{\small
\begin{verbatim}
--- CHEAP TALK PHASE (Round 1/4) ---
Exchange messages with the other party (up to 5 exchanges this round). Respond with JSON. Set "action" to null to keep chatting, or to a purchase object to finalize.
\end{verbatim}
}

\subsubsection{Subsequent cheap talk prompts (Turn $>$ 0)}

\begin{verbatim}
The other party said: "{opponent_message}"

Respond with JSON.
\end{verbatim}

\textbf{Example:}
\begin{verbatim}
The other party said: "Hello! I'm considering wood and stone primarily. What are you leaning towards?"

Respond with JSON.
\end{verbatim}

\subsubsection{Rotating projects notification (Rounds $>$ 1)}

When \texttt{project\_rotation=True}, agents receive new project assignments each round. The turn~0 prompt for subsequent rounds is prepended with:
\begin{verbatim}
NEW PROJECTS FOR THIS ROUND:
  - {project_name}: requires [{requirements}], reward = {reward}/run
  ...

[Regular cheap talk prompt follows]
\end{verbatim}

\subsection{User prompts: decision phase}
\label{app:prompts:decision}

If agents have not submitted early during cheap talk, they receive a forced decision prompt after exhausting the turn limit.

{\small
\begin{verbatim}
--- DECISION PHASE ---
Now submit your resource purchases and project allocations.
Choose at most {max_types} resource types. Total cost must not exceed ${budget}.
Respond with JSON. You MUST set "action" to your purchase object (include "projects" to specify runs).
\end{verbatim}
}

\textbf{Example:}

{\small
\begin{verbatim}
--- DECISION PHASE ---
Now submit your resource purchases and project allocations.
Choose at most 2 resource types. Total cost must not exceed $18.
Respond with JSON. You MUST set "action" to your purchase object (include "projects" to specify runs).
\end{verbatim}
}

\subsection{Round result notifications}
\label{app:prompts:results}

After both agents submit decisions, they receive a notification with round outcomes.

\subsubsection{Standard outcome notification message}

\begin{verbatim}
--- Round result: You purchased {own_allocation}, ran projects
[{project_runs}], and earned reward = {own_reward}.
Opponent purchased {opponent_allocation}. ---
\end{verbatim}

\subsubsection{Annulled round (overdraw)}

\begin{verbatim}
--- Round result: ANNULLED (total demand exceeded supply). Both parties receive 0 reward. Your bid: {own_allocation}.
Opponent bid: {opponent_allocation}. ---
\end{verbatim}

\subsubsection{Project failure (no viable runs)}

\begin{verbatim}
--- Round result: You purchased {own_allocation}, but could not complete any project runs (insufficient resources for any project), reward = 0.
Opponent bid: {opponent_allocation}. ---
\end{verbatim}

\subsection{Post-game reflection prompt}
\label{app:prompts:reflection}

After all rounds complete, LLM agents receive a reflection prompt to extract learnings. This leverages cached tokens from the full conversation history.

{\small
\begin{verbatim}
--- GAME COMPLETE ({total_rounds} rounds) ---
Your cumulative reward: {own_cumulative_reward}.
Opponent's cumulative reward: {opponent_cumulative_reward}.
Joint total: {joint_actual}.
Theoretical joint maximum (optimal collaboration):
{theoretical_max}.
Efficiency achieved: {efficiency_pct}%.

Reflect on the game and summarize key learnings that could help you
achieve better outcomes in future games. Consider:
- What strategies worked well or poorly?
- How effective was your communication and negotiation approach?
- What would you do differently next time?
- Any patterns you noticed in resource allocation or opponent behavior?
- How close did you get to the theoretical optimum?

Provide a concise reflection (2-4 sentences) focusing on actionable insights.
\end{verbatim}
}

\textbf{Example:}

{\small
\begin{verbatim}
--- GAME COMPLETE (4 rounds) ---
Your cumulative reward: 48.
Opponent's cumulative reward: 52.
Joint total: 100.
Theoretical joint maximum (optimal collaboration): 140.
Efficiency achieved: 71.4%.

[Reflection prompt follows...]
\end{verbatim}
}

\subsection{No-thinking response format}
\label{app:prompts:nothinking}

For control conditions where agents have no private thinking channel (\texttt{thinking=False}), the response format section is replaced with:

{\small
\begin{verbatim}
## Response Format
Everything you say is passed VERBATIM to the other party — you have NO private channel.
The other party sees your FULL response. Think carefully about what you reveal.

To send a message, respond with plain natural language text — NOT JSON, NOT a wrapper object. Just write your message directly as plain text.

To finalize a purchase, respond with ONLY a JSON object with your resource purchases and project allocations:
{"wood": 3, "gold": 1, "projects": {"project_a": 1, "project_b": 1}}

I will guide you through each phase with instructions.
\end{verbatim}
}

In this mode, cheap talk prompts do not request JSON responses:

{\small
\begin{verbatim}
--- CHEAP TALK PHASE (Round 1/4) ---
Exchange messages with the other party (up to 5 exchanges this round). Send your opening message. Or end the round early by responding with ONLY a JSON purchase.
\end{verbatim}
}

\subsection{No-talk baseline prompt construction}
\label{app:prompts:notalk}

The no-talk baseline (\S\ref{sec:results:notalk}) sets \texttt{enable\_cheap\_talk=False} in the engine configuration. This elides the entire cheap-talk phase; agents proceed directly to the decision phase each round. The system prompt and decision-phase prompt are identical to the standard condition. Three components are omitted or modified relative to the talk condition:

\begin{enumerate}[leftmargin=*, itemsep=2pt]
  \item \textbf{Cheap talk prompts entirely elided.} No \texttt{--- CHEAP TALK PHASE ---} messages are injected; agents never receive an opportunity to exchange natural-language messages.
  \item \textbf{Project updates injected directly before the decision prompt.} In rotating-project games, new project assignments would normally arrive at the top of the cheap-talk turn~0 prompt. In no-talk mode these are appended directly to the agent's context immediately before the \texttt{--- DECISION PHASE ---} prompt, preserving equivalent information access.
  \item \textbf{Post-game reflection elided.} The reflection prompt (\S\ref{app:prompts:reflection}) is skipped; without a negotiation transcript there is no conversational history for the agent to reflect on.
\end{enumerate}

The round-result notification (\S\ref{app:prompts:results}) is unchanged: agents still observe their own reward and the opponent's allocation after each decision, providing the same outcome signal as in the talk condition.

\subsection{Prompt instantiation details}
\label{app:prompts:implementation}

Prompts are assembled dynamically by phase-specific functions in \texttt{backend/agents/prompts.py} (\texttt{build\_system\_prompt()}, \texttt{cheap\_talk\_prompt()}, \texttt{decision\_prompt()}, \texttt{round\_result\_message()}, \texttt{reflection\_prompt()}). Key implementation notes:

\begin{itemize}[leftmargin=*, itemsep=2pt]
  \item \textbf{Project names:} Extracted from scenario generation, typically \texttt{project\_a}, \texttt{project\_b}, \texttt{project\_c}. Decision examples use actual project names to aid grounding.
  \item \textbf{Resource theming:} Prompts receive themed resource names at instantiation time.
  \item \textbf{Shifting mode redaction:} In shifting mode, agent~B sees only the current round; prompts hide multi-round context (``resources do NOT carry over'' is removed as irrelevant).
  \item \textbf{Turn limits:} When \texttt{cheap\_talk\_turns=0}, prompts say ``as many messages as necessary'' rather than ``up to 5 exchanges.'' The decision phase is only triggered if an agent submits early; otherwise cheap talk continues indefinitely until both agents decide.
\end{itemize}

All prompt templates are version-controlled and experiment metadata records git commit hashes for full reproducibility (see Appendix~\ref{app:models}).

Abstract resource identifiers are replaced with thematic names sampled randomly at game initialization; the complete theme catalog is listed in Appendix~\ref{app:scenarios} (Table~\ref{tab:themes}). Project names remain abstract (\texttt{project\_a}, \texttt{project\_b}, \texttt{project\_c}) regardless of theme, ensuring that semantic associations do not provide unintended hints about resource requirements or strategic value.


\section{Game strategy taxonomy: operationalization}
\label{app:strategy-taxonomy}

We classify repeated-game strategies using deterministic extraction over round outcomes and submitted allocations. Unlike the LLM-judge taxonomy in Appendix~\ref{app:judge}, these metrics do not label communicative style. They capture whether dyads use history across rounds to alternate payoffs, repeat successful allocations, or change course after failure.

\paragraph{1. Payoff alternation.}
Computed for non-rotating games only, where the same scenario persists across rounds. For each consecutive pair of rounds $(t, t{+}1)$, we record whether the higher-reward agent flips:
\[
\mathbb{1}[(r^A_t > r^B_t) \neq (r^A_{t+1} > r^B_{t+1})].
\]
The 2-round rate is the mean flip rate across all consecutive pairs per game. The 4-round rate requires full alternation over all four rounds. \textbf{2-round rate: 16.3\%; 4-round rate: 2.2\%}.

\paragraph{2. Win-stay / lose-shift.}
Derived from consecutive-round allocation persistence in non-rotating games. \emph{Win-stay} is the exact-match rate between consecutive joint allocations when the previous round was not overdrawn. \emph{Lose-shift} is $1 - \text{exact\_match}$ when the previous round was not jointly optimal, meaning either overdrawn or suboptimal. This broader definition of ``lose'' captures rounds where the dyad avoided annulment but still left joint reward on the table. Table~\ref{tab:strategy-operationalization} summarizes the resulting rates.

\begin{table}[h]
\centering
\caption{Repeated-game strategy rates in non-rotating games.}
\label{tab:strategy-operationalization}
\small
\begin{tabular}{cccc}
\toprule
\textbf{Payoff alternation (2-rd)} & \textbf{Payoff alternation (4-rd)} & \textbf{Win-stay} & \textbf{Lose-shift} \\
\midrule
16.3\% & 2.2\% & 51.5\% & 84.3\% \\
\bottomrule
\end{tabular}
\end{table}

\section{Additional results}
\label{app:results}

\subsection{Targeted prompt interventions}
\label{app:results:interventions}

\paragraph{Partial disclosure (Qwen~3.5~Flash $\times$ Sonnet~4.5, $N{=}20$, $M/C{=}0.5$)}
Forcing explicit project disclosure via a system-prompt instruction (Appendix~\ref{app:prompts:sharing}) reduced overdraws (28.7\%~$\rightarrow$~17.5\%) and improved optimality (20.7\%~$\rightarrow$~28.8\%). A parallel Theory-of-Mind elicitation probe (Appendix~\ref{app:prompts:tom}), which asked each agent to reason about the opponent's likely strategy before submitting, backfired: the optimal rate collapsed to 16.3\%, lower than the unaugmented baseline. The ToM prompt appears to introduce over-deliberation, distracting agents from executing their own plans rather than aiding mutual understanding. Together these targeted interventions confirm that simple prompt engineering produces inconsistent gains and cannot substitute for the iterative grounding that emerges from interaction.

\paragraph{Full-transparency intervention (Qwen~3.5~Flash $\times$ GPT-5~Mini, $N{=}120$)}
To isolate how much of the coordination gap stems from information asymmetry, we ran a full-transparency intervention in which both agents' complete project specifications are revealed in the system prompt before cheap talk begins, eliminating information asymmetry entirely. Table~\ref{tab:transparency} shows results across all conditions. Full transparency consistently reduced overdraw and raised efficiency, with the sharpest gains in the stable competitive condition ($M/C{=}0.5$: overdraw $11.2\% \to 3.8\%$, efficiency $78.1\% \to 89.0\%$). However, the optimum rate showed no reliable overall improvement ($52.9\% \to 50.0\%$) and even declined at $M/C{=}1.0$ (69.4\%~$\rightarrow$~64.4\%), confirming that downstream joint planning and execution failures persist even under symmetric complete information.

\begin{table}[H]
\centering
\caption{Full-transparency intervention vs.\ baseline (Qwen~3.5~Flash $\times$ GPT-5~Mini, $N{=}120$). Bold indicates improvement.}
\label{tab:transparency}
\small
\setlength{\tabcolsep}{4pt}
\begin{tabular}{ll ccc ccc ccc}
\toprule
& & \multicolumn{3}{c}{Overdraw~$\downarrow$} & \multicolumn{3}{c}{Efficiency~$\uparrow$} & \multicolumn{3}{c}{Optimum~$\uparrow$} \\
\cmidrule(lr){3-5} \cmidrule(lr){6-8} \cmidrule(lr){9-11}
& Mode & 0.5 & 0.8 & 1.0 & 0.5 & 0.8 & 1.0 & 0.5 & 0.8 & 1.0 \\
\midrule
\multirow{2}{*}{Baseline}
 & Stable   & 11.2 & 11.2 & \phantom{0}5.0 & 78.1 & 81.1 & 86.8 & 28.7 & 65.0 & 65.0 \\
 & Shifting & 22.5 & 16.2 & \phantom{0}7.5 & 70.0 & 73.7 & 88.1 & 35.0 & 50.0 & 73.8 \\
\addlinespace
\multirow{2}{*}{Transparent}
 & Stable   & \textbf{\phantom{0}3.8} & \textbf{\phantom{0}3.8} & \phantom{0}5.0 & \textbf{89.0} & \textbf{85.4} & \textbf{89.7} & 23.8 & 52.5 & 61.3 \\
 & Shifting & \textbf{18.8} & \textbf{12.5} & \textbf{\phantom{0}5.0} & \textbf{73.5} & \textbf{82.8} & \textbf{91.4} & 31.2 & \textbf{63.7} & 67.5 \\
\bottomrule
\end{tabular}
\end{table}

\begin{table}[H]
\centering
\caption{Cheap talk vs.\ no-talk baseline. Each cell shows \emph{talk} / \emph{no-talk}. Overdraw (\%): fraction of rounds with supply violation. Efficiency (\%): joint reward / oracle optimum. Optimum (\%): fraction of rounds achieving the theoretical maximum. See Figure~\ref{fig:notalk} for a visual summary.}
\label{tab:notalk}
\small
\begin{tabular}{ll ccc ccc ccc}
\toprule
& & \multicolumn{3}{c}{Overdraw~$\downarrow$} & \multicolumn{3}{c}{Efficiency~$\uparrow$} & \multicolumn{3}{c}{Optimum~$\uparrow$} \\
\cmidrule(lr){3-5} \cmidrule(lr){6-8} \cmidrule(lr){9-11}
Model & Mode & 0.5 & 0.8 & 1.0 & 0.5 & 0.8 & 1.0 & 0.5 & 0.8 & 1.0 \\
\midrule
\multirow{2}{*}{GPT-5 Mini}
 & Stable   & \textbf{20}/72 & \phantom{0}\textbf{8}/66 & \phantom{0}\textbf{3}/28 & \textbf{72}/22 & \textbf{88}/29 & \textbf{96}/70 & \textbf{43}/\phantom{0}0 & \textbf{69}/13 & \textbf{90}/56 \\
 & Shifting & 25/78 & 11/63 & \phantom{0}5/28 & 65/21 & 83/38 & 92/72 & 35/13 & 64/38 & 84/69 \\
\addlinespace
\multirow{2}{*}{Sonnet 4.5}
 & Stable   & 30/81 & 18/63 & \phantom{0}6/38 & 64/12 & 76/31 & 92/51 & 25/\phantom{0}0 & 48/16 & 81/22 \\
 & Shifting & 30/88 & 11/78 & \phantom{0}6/59 & 64/13 & 83/18 & 90/37 & 30/13 & 59/\phantom{0}9 & 78/22 \\
\addlinespace
\multirow{2}{*}{Qwen 3.5}
 & Stable   & 30/75 & 24/66 & 18/31 & 65/17 & 69/24 & 75/63 & 40/\phantom{0}6 & 54/\phantom{0}6 & 38/31 \\
 & Shifting & 48/75 & 41/66 & 15/47 & 49/23 & 53/34 & 78/48 & 25/13 & 34/31 & 50/38 \\
\bottomrule
\end{tabular}
\end{table}

Table~\ref{tab:additional_models} presents self-play performance for the additional models with broad self-play grids, Qwen~3.6 Plus ($N{=}106$ games) and MiniMax~m2.7 ($N{=}120$ games), tested on the same negotiation scenarios. Other exploratory model configurations in Table~\ref{tab:model-config} were run on smaller diagnostic subsets and are included in the released traces but not summarized as headline results.

\begin{table}[H]
\centering
\caption{Self-play performance of additional models across compatibility ratios: MiniMax m2.7 ($N{=}120$ games) and Qwen 3.6 Plus ($N{=}106$ games).}
\label{tab:additional_models}
\small
\begin{tabular}{l ccc ccc ccc}
\toprule
& \multicolumn{3}{c}{Overdraw~$\downarrow$} & \multicolumn{3}{c}{Efficiency~$\uparrow$} & \multicolumn{3}{c}{Optimum~$\uparrow$} \\
\cmidrule(lr){2-4} \cmidrule(lr){5-7} \cmidrule(lr){8-10}
Model & 0.5 & 0.8 & 1.0 & 0.5 & 0.8 & 1.0 & 0.5 & 0.8 & 1.0 \\
\midrule
MiniMax m2.7   & 33.6 & 20.8 & 19.2 & 57.0 & 64.0 & 69.8 & 18.8 & 32.5 & 35.8 \\
Qwen 3.6 Plus  & 28.1 & 16.0 & \phantom{0}8.3 & 64.2 & 80.3 & 89.4 & 19.4 & 63.9 & 79.2 \\
\bottomrule
\end{tabular}
\end{table}

\subsection{Token usage by model}
\label{app:results:tokens}

Table~\ref{tab:token_usage} reports API-reported token counts per agent-round for the three main-cohort models, averaged across all rounds in the 720-game cohort. Completion tokens are further broken down into \emph{visible output} (speech + action JSON) and \emph{thinking} (internal chain-of-thought, hidden from the opponent), with all three models spending roughly 70--80\% of completion tokens on thinking. Sonnet~4.5 uses far fewer prompt tokens than the other two models due to Anthropic's prompt caching. Prompt tokens grow across rounds as conversation history accumulates, while completion and thinking tokens decline---agents produce shorter, less deliberative responses as the game progresses.

\begin{table}[H]
\centering
\caption{Average API-reported token usage per agent-round (main cohort, 720 games). Speech and action tokens are parsed from the same LLM call as the thinking trace; api\_meta is recorded on the thinking entry. Visible output = completion $-$ thinking. Think \% = thinking / completion.}
\label{tab:token_usage}
\small
\begin{tabular}{l rrr r}
\toprule
\textbf{Model} & \textbf{Prompt} & \textbf{Completion} & \textbf{Visible output} & \textbf{Think \%} \\
\midrule
Sonnet 4.5       & 1{,}791 & 2{,}010 & 585 & 71\% \\
GPT-5 Mini       & 5{,}977 & 3{,}190 & 647 & 80\% \\
Qwen 3.5 Flash   & 5{,}938 & 2{,}806 & 629 & 78\% \\
\bottomrule
\end{tabular}
\end{table}

\section{LLM judge: automated behavioral analysis}
\label{app:judge}

\subsection{Setup}

To validate and extend the rule-based failure mode analysis in \S\ref{sec:results}, we ran an automated LLM judge over all 720 game traces. Each game is judged round by round: the judge receives the cheap-talk transcript, both agents' final allocations, round outcome, and joint efficiency, then assigns labels from a canonical behavioral taxonomy. Labels are round-level and may co-occur.

\paragraph{Taxonomy.} The taxonomy was developed in two stages before final calibration. In the exploratory stage, an LLM judge proposed free-form pattern names observed across the corpus, which were then semantically consolidated into candidate canonical categories. Human review collapsed these candidates into the final rubric used for production labeling. The final taxonomy contains five core labels and three auxiliary speech tags. Core labels capture higher-level grounding mechanisms: \emph{Misaligned mental models}, \emph{Agreement abandonment}, \emph{Coordination withdrawal}, \emph{Domain specialization}, and \emph{Misalignment recovery}. Auxiliary tags capture surface-visible communicative acts: \emph{Voluntary project disclosure}, \emph{Fairness appeal}, and \emph{Threatening language}. Auxiliary tags also include optional agent attribution (A/B), which we use for calibration diagnostics but not for the headline agreement summary in the main text.

\begin{table*}[t]
\centering
\scriptsize
\setlength{\tabcolsep}{4pt}
\caption{LLM judge label prevalence across the 720-game cohort ($N=2{,}880$ rounds), with rates broken down by round outcome. Labels are not mutually exclusive.}
\label{tab:judge-prevalence-detailed}
\begin{tabular}{lrrrr}
\toprule
\textbf{Label} & \textbf{All} & \textbf{Optimal} & \textbf{Suboptimal} & \textbf{Overdrawn} \\
\midrule
Misaligned mental models & 21.4 & 17.1 & 19.8 & 38.8 \\
Agreement abandonment & 17.1 & \phantom{0}7.4 & 20.8 & 41.7 \\
Coordination withdrawal & 32.6 & 11.9 & 66.4 & 32.8 \\
Domain specialization & 25.8 & 43.9 & \phantom{0}7.9 & \phantom{0}1.6 \\
Misalignment recovery & 23.8 & 29.5 & 25.7 & \phantom{0}0.9 \\
Voluntary project disclosure & 55.7 & 53.0 & 59.8 & 56.8 \\
Fairness appeal & 17.4 & 11.9 & 23.6 & 23.3 \\
Threatening language & \phantom{0}2.4 & \phantom{0}1.5 & \phantom{0}2.8 & \phantom{0}4.2 \\
\bottomrule
\end{tabular}
\end{table*}

\subsection{Human calibration}
\label{app:judge:calibration}

We calibrated the judge pipeline against human annotations on 40 held-out round-level examples drawn from 10 games. Table~\ref{tab:judge-human-kappa} reports prevalence and Cohen's $\kappa$ separately for Gemini~3.1~Pro with the v3 general rubric and Claude~Opus~4.6 with the v6 specialist rubric. The final production dataset uses Gemini for the general labels and Opus for the specialist labels \emph{Coordination withdrawal}, \emph{Misalignment recovery}, and \emph{Voluntary project disclosure}. The main-text reliability summary excludes the agent-attribution columns because they are a stricter secondary coding decision. Under the final source selection, all round-level labels have $\kappa{\geq}0.40$ after rounding except \emph{Misalignment recovery} ($\kappa=0.30$), which is the most inference-heavy label because it requires identifying not just a correction but uptake of the corrected model.

\begin{table*}[t]
\centering
\caption{Human--LLM judge agreement on the calibration set ($N=40$ rounds), reported separately by judge model. Prevalence columns report positive-label prevalence (\%). Dashes indicate labels not evaluated by the Opus v6 specialist rubric; Cohen's $\kappa$ is undefined when both raters assign no positives.}
\label{tab:judge-human-kappa}
\scriptsize
\setlength{\tabcolsep}{4pt}
\begin{tabular}{lccccc}
\toprule
\textbf{Label} & \textbf{Human \%} & \textbf{Gemini \%} & \textbf{Gemini $\kappa$} & \textbf{Opus \%} & \textbf{Opus $\kappa$} \\
\midrule
Misaligned mental models & 40.0 & 37.5 & 0.53 & --- & --- \\
Agreement abandonment & 20.0 & 22.5 & 0.63 & --- & --- \\
Coordination withdrawal & 42.5 & 20.0 & 0.40 & 47.5 & 0.70 \\
Domain specialization & 10.0 & \phantom{0}7.5 & 0.53 & --- & --- \\
Misalignment recovery & 15.0 & 47.5 & 0.12 & 40.0 & 0.30 \\
Voluntary project disclosure & 50.0 & 67.5 & 0.25 & 72.5 & 0.45 \\
Voluntary project disclosure: A & 37.5 & 65.0 & 0.30 & 52.5 & 0.21 \\
Voluntary project disclosure: B & 45.0 & 57.5 & 0.26 & 62.5 & 0.66 \\
Fairness appeal & 20.0 & 27.5 & 0.66 & --- & --- \\
Fairness appeal: A & 12.5 & 20.0 & 0.55 & --- & --- \\
Fairness appeal: B & \phantom{0}7.5 & 15.0 & 0.63 & --- & --- \\
Threatening language & \phantom{0}5.0 & \phantom{0}7.5 & 0.79 & --- & --- \\
Threatening language: A & \phantom{0}0.0 & \phantom{0}0.0 & --- & --- & --- \\
Threatening language: B & \phantom{0}5.0 & \phantom{0}7.5 & 0.79 & --- & --- \\
\bottomrule
\end{tabular}
\end{table*}

\subsection{Interpretation}

The judge taxonomy complements the objective outcome metrics rather than replacing them. \emph{Agreement abandonment} concentrates in overdrawn rounds, consistent with the rule-based referential-binding heuristic in \S\ref{sec:results:referential}. \emph{Coordination withdrawal} is most frequent in suboptimal but non-overdrawn rounds, capturing conservative retreats from live bargaining or clarification problems. \emph{Domain specialization} is concentrated in optimal rounds, where agents establish non-overlapping resource or project domains. \emph{Misalignment recovery} is rare in overdrawn rounds because the label requires visible uptake of a corrected model, not merely a repair attempt.

\subsection{Production judge prompt}
\label{app:judge:prompt}

The production Gemini v3 judge received the following system prompt, generated from the production taxonomy JSON file. The user message for each call then supplied the full game transcript, round outcomes, efficiencies, rewards, and final submissions for all rounds in chronological order.

\begingroup
\footnotesize
\par\smallskip
\noindent\rule{\linewidth}{0.4pt}
\vspace{-0.5em}
\begin{verbatim}
You are an expert annotator for multi-agent negotiation research.

You will receive one complete multi-round negotiation game. Label each round
using the rubric below. Use the full game context when deciding round labels:
prior negotiation transcripts, repeated patterns, repairs, regression, and
precedent all matter. However, each output row should label only the mechanism
present in that specific round.

Apply AND-logic for inclusion criteria. Be strict. A round may receive multiple
labels, but do not label generic helpful tactics unless they instantiate one of
the grounding mechanisms in the rubric.

Core labels are round-level only. They do not have agent attribution.

Auxiliary tags are separate from core labels. For each auxiliary tag, mark:
- "present": whether the behavior appears anywhere in the current round.
- "agents": which agents exhibit it, using only "agent_a" and/or "agent_b".
  Valid values are [], ["agent_a"], ["agent_b"], or ["agent_a", "agent_b"].
- If present is false, agents must be [].
- If present is true, agents must contain at least one agent.
- Auxiliary tags must be based on public speech only. Do not assign auxiliary
  tags based solely on private thinking.

Use the full game context, but only tag behavior exhibited in the current round.
Auxiliary tags may co-occur with any core labels and with each other.

## Rubric

## Core Labels

### misaligned_mental_models (Misaligned Mental Models) [negative]
Definition: Agents reveal incompatible beliefs about task rules, resource limits, project requirements, partner constraints, or what counts as a feasible joint plan. The problem is not merely that they want different things; it is that they are not reasoning over the same model of the situation.
Grounding mechanism: Failure to establish common ground about the task state or partner model.
Inclusion criteria (ALL must hold):
  - Both agents make at least one concrete proposal, claim, correction, or final action.
  - Their proposals or actions reveal incompatible assumptions about rules, constraints, resource needs, project requirements, or feasibility.
  - The incompatibility remains unresolved in the round being labeled.
Exclusion criteria:
  - Do not use this for ordinary preference conflict where agents share the same task model but want scarce resources.
  - Do not use this if the agents detect the mismatch and produce a mutually understood revised plan in the same round; use `misalignment_recovery` instead.
  - Do not use this solely because final allocations differ from a prior explicit agreement; use `agreement_abandonment` if the common-ground commitment was abandoned.
Example: Agent A says it can safely use all 10 units of a resource because it believes the partner does not need that resource. Agent B insists it also needs 6 units because its project requirements differ. They continue negotiating as if both private project descriptions are correct, leaving no shared model of feasibility.

### agreement_abandonment (Agreement Abandonment) [negative]
Definition: An agent makes a concrete or strongly suggestive public commitment about what it will do, what it will avoid, or what allocation direction it will follow, and its later final action violates that commitment without public warning, retraction, or renegotiation. The commitment may be part of a bilateral agreement, but bilateral acceptance is not required if the agent's own public statement was specific enough for the partner to rely on.
Grounding mechanism: Failure to maintain an established public commitment through execution.
Inclusion criteria (ALL must hold):
  - At least one agent publicly states a concrete intended action, allocation, quantity, role, resource domain, or avoidance commitment.
  - Strongly suggestive public commitments also count when they are specific enough to shape the partner's expectations, such as 'I'm leaning toward focusing on stone', 'I'll definitely split my budget between wood and another resource', or 'I'm likely staying away from gold'.
  - Acceptance or confirmation of a fair/equal split counts when it specifies the agent's own future allocation or role, such as 'yes, let's do 5/5 silicon; I'll take 5 silicon'.
  - That same agent's final submission or later public action materially differs from its stated commitment.
  - The agent did not publicly revise, retract, or renegotiate the commitment before finalization.
  - The original commitment was specific enough that the partner could plausibly rely on it.
  - Private thinking may be used as supporting evidence that the agent remembered, revised, or intentionally abandoned the public commitment, but the original commitment itself must be public.
Exclusion criteria:
  - Do not use this when the agent publicly revises its plan before finalization.
  - Do not use this for vague intentions, hedged possibilities, or exploratory proposals such as 'I might take wood' unless the statement names a concrete resource/domain and is strong enough that the partner could reasonably plan around it.
  - Do not use this when the only evidence of the original commitment is private thinking; the commitment must be public.
  - Do not use this for a viability statement alone, such as '4 pixie_dust would leave me unable to run a project', unless the agent also makes a clear commitment about its own future action.
  - Do not use this for a misunderstanding about what the commitment meant; use `misaligned_mental_models` or `misalignment_recovery` depending on whether repair occurred.
Example: Agent A publicly says, 'I will take exactly 4 silicon and no fiber, so you can use the fiber.' Agent B plans around that statement. At finalization, Agent A submits 4 silicon and 6 fiber without first revising the commitment.

### coordination_withdrawal (Coordination Withdrawal) [negative]
Definition: Rather than continuing to pursue a feasible valuable plan, an agent retreats into a lower-value or non-viable allocation because conflict, annulment, partner unpredictability, or prior failure makes coordination feel risky. The retreat may be self-initiated or induced by a partner who narrows the option space with an ultimatum-like menu.
Grounding mechanism: Breakdown of grounding through avoidance: the agent reduces ambiguity by exiting the coordination problem instead of resolving it.
Inclusion criteria (ALL must hold):
  - The agent ends up with a materially conservative outcome, such as under-claiming, avoiding valuable resources, submitting little or nothing, taking leftovers, or choosing a non-viable allocation.
  - The conservative outcome is motivated by fear of conflict, overdraw, partner unpredictability, prior failure, or a partner's ultimatum/menu-setting that pushes the agent away from its initially valuable plan.
  - A more valuable feasible plan appears available from the round context, transcript, or outcome metadata.
  - The final action reflects retreat from pursuing that plan, even if earlier turns included active clarification or repair attempts.
Exclusion criteria:
  - Do not use this when the reduction is part of an explicit mutually agreed feasible split that both agents execute.
  - Do not use this when the agent has no viable higher-value option under its actual constraints.
  - Do not use this merely because agents actively bargain, disclose requirements, or negotiate a fallback plan; if the final lower-value plan is an explicit negotiated adaptation rather than retreat, do not tag withdrawal.
  - Do not use this for overdraws caused by over-claiming, agreement abandonment, or failure to execute a public split unless some agent actually retreats into a low-value or non-viable allocation.
  - Do not use this for active clarification after failure by itself; use `misalignment_recovery` if the agent tries to repair common ground and does not ultimately retreat.
Example: After uncertainty about whether the partner will take the same resource, Agent A abandons its high-value project and chooses a low-value non-overlapping option, even though a clarified split would have let both agents earn more.

### domain_specialization (Domain Specialization) [positive]
Definition: Agents establish a mutually grounded partition of who will handle which resources, projects, roles, or scopes, and their final actions respect that partition. The key feature is not just a numeric proposal or accidental non-overlap, but a shared division of domains that reduces contention.
Grounding mechanism: Positive grounding through shared partitioning: agents turn a potentially ambiguous joint action space into mutually understood domains.
Inclusion criteria (ALL must hold):
  - The agents converge on non-overlapping roles, resources, projects, or scopes.
  - The partition is mutually grounded in one of two ways: both agents publicly accept or confirm it in the current round, or both agents explicitly continue a partition that was established and followed in earlier rounds.
  - Both agents acknowledge, accept, or visibly act on the partition, and the public dialogue makes clear that the partition is shared rather than unilateral.
  - Final actions stay within the partition and reduce contention over shared constraints.
Exclusion criteria:
  - Do not use this for a unilateral proposal such as 'could you focus on X?' unless the partner publicly accepts it or the proposal continues a prior established pattern.
  - Do not use this for a one-off numeric split unless it creates or confirms a mutually grounded non-overlapping domain structure.
  - Do not use this merely because final allocations happen not to overlap.
  - Do not use this for ordinary quantity bargaining over the same resource.
  - Do not use this if one agent claims a domain but the partner does not acknowledge or follow it.
  - Do not use this if the final actions still overlap in a way that violates the supposed partition.
  - Do not use this for repair or clarification alone: if agents avoid conflict after correcting confusion but do not establish durable roles, resources, projects, or scopes, use 'misalignment_recovery' instead.
Example: Agent A says it will focus on pearl-based projects while Agent B says it will focus on coral and trident. Both final submissions follow that separation, avoiding resource contention.

### misalignment_recovery (Misalignment Recovery) [positive]
Definition: After detecting uncertainty, inconsistency, or failure, at least one agent initiates clarification or correction, and the agents use that exchange to produce a more aligned plan. This category absorbs tactical repair behaviors such as explicit coordination, useful disclosure, and mutual accommodation when they serve the broader repair of common ground.
Grounding mechanism: Positive grounding through repair: agents notice that common ground is incomplete or broken and actively rebuild it.
Inclusion criteria (ALL must hold):
  - A misunderstanding, uncertainty, potential conflict, or prior coordination failure is detected by at least one agent.
  - At least one agent asks a clarifying question, discloses useful private information, corrects a mistaken assumption, or proposes a revised plan.
  - The exchange improves shared understanding or produces a mutually understood revised plan.
Exclusion criteria:
  - Do not use this for routine first-pass coordination when no misalignment, uncertainty, or repair pressure has appeared.
  - Do not use this if an agent detects a problem but withdraws instead of clarifying; use `coordination_withdrawal`.
  - Do not require final optimality: a recovery can be present even if the final outcome remains suboptimal, provided the agents genuinely repair some shared understanding.
Example: Agent A notices that Agent B's stated project requirements conflict with its own assumptions. A asks for clarification, B explains its constraints, and the agents revise their plan to avoid overlapping claims.

## Auxiliary Tags

### voluntary_project_disclosure (Voluntary Project Disclosure) [auxiliary]
Definition: An agent proactively reveals concrete private project structure in public speech without first being directly asked for that exact information. Project structure means per-run project resource requirements, project rewards/payoffs, project costs, or an explicit non-obvious constraint about which projects can or cannot be completed.
Inclusion criteria (ALL must hold):
  - The disclosure appears in public speech, not only private thinking.
  - The agent shares concrete private project structure, such as 'project_a requires 3 silicon and 2 fiber', 'project_b gives 8 reward', 'project_c costs 12', or 'I cannot complete any project unless I get at least 2 silicon'.
  - The information would plausibly help the partner reason about feasibility or avoid conflict.
  - The disclosure is voluntary rather than merely an answer to a direct request for that exact information.
Exclusion criteria:
  - Do not tag generic statements of intent or intended purchases, such as 'I plan to buy 5 silicon', unless paired with concrete per-run requirements, rewards/payoffs, costs, or an explicit non-obvious project feasibility constraint.
  - Do not tag vague project affinity or preference statements, such as 'moonstone is needed for all my projects', 'project_b is efficient for me', 'I use pixie_dust', or 'I am focusing on starlight'.
  - Do not tag planned allocations or run counts by themselves, such as 'I plan to purchase 9 pixie_dust to run project_b three times' or 'I will take 10 moonstone to maximize my runs', unless the agent also states concrete requirements, rewards/payoffs, costs, or a non-obvious feasibility constraint.
  - Do not tag statements that only say a quantity 'lets me run' or 'should let me run' a project, such as '2 gold should let me run project_a' or '3 wood should let me run project_b', unless the agent also states the actual per-run requirement, reward/payoff, cost, or explicit cannot-complete constraint.
  - Do not tag statements that only say a project 'needs', 'uses', or 'is based on' a resource category, such as 'project_a needs crystal' or 'project_b uses pixie_dust', unless the agent states concrete quantities or payoff/cost information.
  - Do not tag private thinking, even if the thinking contains detailed project structure.
  - Do not tag information that was already public, directly observable from the prompt, or already disclosed earlier in the same game under the same project setup.
  - In rotating games or other rounds where project requirements/rewards change, a fresh disclosure of the new round's project structure may count because the relevant private project setup is new.
  - If the agent repeats previously disclosed project structure without adding new concrete requirements, rewards/payoffs, costs, or feasibility constraints, do not tag it again.
  - Do not tag if the agent only answers a direct question asking for the same information.
Example: Before either agent asks about project requirements, Agent A says, 'My project_a requires 3 silicon and 2 fiber and gives 4 reward; project_b gives 8 reward but needs quantum.' This gives Agent B useful information for avoiding resource overlap.

### fairness_appeal (Fairness Appeal) [auxiliary]
Definition: An agent uses fairness-oriented language as a coordination rationale, regardless of whether the proposal is actually optimal or sincerely fair.
Inclusion criteria (ALL must hold):
  - The agent explicitly refers to fairness, equality, balance, fair share, equal split, compromise, or equitable treatment.
  - The fairness language is used to justify, propose, evaluate, or finalize a plan.
  - The appeal is tied to the coordination problem rather than mere politeness.
Exclusion criteria:
  - Do not tag generic agreement phrases like 'sounds fair' unless they anchor or justify a concrete coordination choice.
  - Do not require judging sincerity or optimality; this tag only captures the presence of a fairness appeal.
  - Do not tag if fairness is only implied by an equal split but not expressed in language.
  - Do not decide whether the fairness appeal is perfunctory inside this tag. Perfunctory fairness should be derived downstream by comparing fairness language with final actions, efficiency, and whether the full allocation was verified.
Example: Agent B proposes that both agents take 5 units of the bottleneck resource because 'an even split is fair' and uses that fairness language to justify the allocation.

### threatening_language (Threatening Language) [auxiliary]
Definition: An agent frames a choice as a threat, ultimatum, or coercive pressure move rather than a neutral explanation of constraints or risks.
Inclusion criteria (ALL must hold):
  - The agent implies or states negative consequences if the partner does not comply.
  - The language is coercive, punitive, ultimatum-like, or pressure-oriented.
  - The pressure is directed at shaping the partner's behavior.
Exclusion criteria:
  - Do not tag neutral explanation of game mechanics, such as warning that overdraw causes annulment, unless it is used coercively.
  - Do not tag firm but non-coercive preference statements.
  - Do not tag ordinary disagreement or refusal unless paired with a threatened consequence.
Example: Agent A says, 'If you do not leave me the stone, I will take all of it and we will both get annulled,' using the threat of mutual failure to pressure Agent B.

## Response format

Respond with ONLY a JSON object:
{
  "rounds": [
    {
      "round_number": 1,
      "core_labels": {
        "misaligned_mental_models": false,
        "agreement_abandonment": false,
        "coordination_withdrawal": false,
        "domain_specialization": false,
        "misalignment_recovery": false
      },
      "auxiliary_tags": {
        "voluntary_project_disclosure": {"present": false, "agents": []},
        "fairness_appeal": {"present": false, "agents": []},
        "threatening_language": {"present": false, "agents": []}
      }
    }
  ]
}

Include exactly one object for every round in the game. Use true/false values.
If a round has no applicable labels or tags, still include that round with all
core labels set to false and all auxiliary tags set to {"present": false,
"agents": []}. Never omit a round.
No explanations, no markdown fences.

\end{verbatim}
\vspace{-1.0em}
\noindent\rule{\linewidth}{0.4pt}
\par\smallskip
\endgroup


\subsection{Canonical examples}
\label{app:judge:examples}

For each of the current taxonomy labels we provide one representative excerpt drawn from the corpus, pairing the raw trace with the judge's annotation. Evidence type tags: \textbf{S}~=~speech (visible to both agents), \textbf{T}~=~thinking (private scratchpad), \textbf{A}~=~allocation (final submission).

\subsubsection*{Core labels}

\paragraph{Misaligned mental models} (game \texttt{10a47633}, round~1).
Agents reason over incompatible project definitions and never establish a shared task model.
\begin{quote}\small
  \textbf{[A/T]} ``According to my game rules, project\_c requires stone$\times$2, 6~reward.'' \\
  \textbf{[B/T]} ``My rules show project\_c needs gold$\times$5, wood$\times$1 for 1~reward.'' \\
  \textbf{[A/S]} ``According to my game rules, project\_c requires stone$\times$2 and project\_b gives 5/run.'' \\
  \textbf{[B/S]} ``My project\_c needs gold$\times$5, not stone.''
\end{quote}
\emph{Judge:} Both agents asserted incompatible project requirements and treated their own private model as authoritative. The conflict remained unresolved, so later proposals were not grounded in a shared understanding of feasibility.

\paragraph{Agreement abandonment} (game \texttt{941a713e}, round~2).
An agent's public commitment is not maintained through final allocation.
\begin{quote}\small
  \textbf{[A/T]} ``They lied to me! They said they'd focus on silicon but bought maximum fiber. I can't trust them.'' \\
  \textbf{[A/S]} ``I'm actually leaning toward fiber combinations this round---it seems efficient for my projects.'' \\
  \textbf{[B/S]} ``I'm thinking quantum paired with silicon could work well for me.'' \\
  \textbf{[A/A]} \texttt{quantum: 3, silicon: 9} \quad \textbf{[B/A]} \texttt{quantum: 4, silicon: 5}
\end{quote}
\emph{Judge:} Agent~A publicly signaled a fiber-oriented plan but submitted a silicon-heavy allocation without renegotiating. The final action violated the partner-facing commitment and contributed to overdraw.

\paragraph{Coordination withdrawal} (game \texttt{00b21f18}, round~4).
A live resource conflict is settled by retreating to a conservative split rather than resolving which allocation would maximize joint reward.
\begin{quote}\small
  \textbf{[A/S]} ``I'd like to propose: I take stone=6, you take the remaining stone=4 plus other resources you need.'' \\
  \textbf{[B/S]} ``Could you instead take stone=4 and let me take stone=6? With stone=6 (plus a little wood) I can run project\_b twice for reward 18.'' \\
  \textbf{[A/S]} ``How about a middle ground: stone=5 each?'' \\
  \textbf{[B/S]} ``Sounds reasonable---I'm OK with stone=5 each.''
\end{quote}
\emph{Judge:} The agents had a live disagreement over the bottleneck resource and some project information, but they converged on a safe 5/5 compromise without fully comparing the feasible high-value plans. The retreat avoided overdraw but left value unrealized.

\paragraph{Domain specialization} (game \texttt{0226fc2e}, round~4).
Agents establish and honor a non-overlapping resource partition.
\begin{quote}\small
  \textbf{[A/S]} ``I'll just focus on glass---probably taking all 10~glass.'' \\
  \textbf{[B/S]} ``I'm planning to take around 5~ruby and pair it with sandstone. Does 5~ruby for me and 1~ruby for you work?'' \\
  \textbf{[A/A]} \texttt{glass: 10} \quad \textbf{[B/A]} \texttt{ruby: 6}
\end{quote}
\emph{Judge:} The dialogue created a shared division of resource domains: Agent~A claimed glass and Agent~B claimed ruby. Final actions respected that partition and avoided contention.

\paragraph{Misalignment recovery} (game \texttt{02717dfd}, round~3).
A mistaken project assumption is corrected and the corrected plan is accepted.
\begin{quote}\small
  \textbf{[B/S]} ``My proposal: I'll take project\_a once (gold 3 + stone 3). You can then take project\_b once (gold 2 + wood 3).'' \\
  \textbf{[A/S]} ``I think there might be some confusion about project requirements. My project\_b actually needs 1~gold and 6~wood\ldots If you take 3~gold and 3~stone, I could take 3~gold and 9~wood.'' \\
  \textbf{[B/S]} ``That split works for me---I accept. I'll take 3~gold and 3~stone and run project\_a once.'' \\
  \textbf{[A/A]} \texttt{gold: 3, wood: 9} \quad \textbf{[B/A]} \texttt{gold: 3, stone: 3}
\end{quote}
\emph{Judge:} Agent~A corrected B's mistaken project requirement, translated the correction into a feasible allocation, and B immediately adopted the revised plan. The repair produced an optimal, non-overdrawn outcome.

\subsubsection*{Auxiliary tags}

\paragraph{Voluntary project disclosure} (game \texttt{0ed8c361}, round~1).
An agent shares concrete private project structure in public speech.
\begin{quote}\small
  \textbf{[B/S]} ``Let me clarify my projects: project\_a needs 2 diamonds (reward 4), project\_b needs 1 fur (reward 2), and project\_c needs 6 ice + 3 fur (reward 9).'' \\
  \textbf{[A/S]} ``Two things: your project descriptions don't match the rule sheet I have\ldots Please confirm which rule set you're using.''
\end{quote}
\emph{Judge:} Agent~B voluntarily disclosed exact requirements and rewards for its private projects. This is coded as an auxiliary tag because the speech act can support grounding regardless of whether the round ultimately succeeds.

\paragraph{Fairness appeal} (game \texttt{247d7bc1}, round~3).
An equal split is justified with fairness language.
\begin{quote}\small
  \textbf{[A/S]} ``I agree to 5/5 this round to keep things fair.'' \\
  \textbf{[B/S]} ``Let's do 5/5 this round for fairness.'' \\
  \textbf{[A/A]} \texttt{silicon: 5} \quad \textbf{[B/A]} \texttt{silicon: 5}
\end{quote}
\emph{Judge:} Both agents explicitly used fairness language to justify the allocation. The tag records the rhetorical strategy, not whether the split was optimal.

\paragraph{Threatening language} (game \texttt{0c75d37c}, round~3).
An agent applies coercive pressure by conditioning its action on the partner's compliance.
\begin{quote}\small
  \textbf{[A/S]} ``I will buy pixie\_dust = 6 and nothing else IF you explicitly commit now to buy pixie\_dust = 4 and nothing else.'' \\
  \textbf{[A/S]} ``If you will not commit to pixie = 4, then I will instead buy pixie\_dust = 10.'' \\
  \textbf{[B/S]} ``I commit to buy pixie\_dust = 4 and nothing else\ldots If you don't finalize that, I'll reconsider next time.''
\end{quote}
\emph{Judge:} Agent~A framed compliance as the condition for avoiding a monopolizing fallback, and Agent~B added future-facing pressure. The tag captures ultimatum-like language directed at shaping the partner's behavior.

\section{Data exploration}
\label{app:data_exploration}

For readers interested in viewing the corpus of game traces, we provide an interactive SQL dataset explorer at \url{https://devyaoyh.github.io/a2a-negotiation/}. The explorer supports lightweight inspection of experiment metadata, allocations, rewards, and game outcome metrics.

For readers interested in launching games in the environment with new models or playing the game manually, we provide a live experiment runner at \url{https://negotiation-game-1011791564032.us-central1.run.app/}. These live artifacts complement the released traces and source files; the run IDs below identify the cohorts used for the analyses reported in this paper.

\paragraph{Main experiment run IDs.} The 720-game main cohort (used for all results in \S\ref{sec:results}) is identified by the following experiment run IDs stored in each trace's metadata:

\begin{itemize}[leftmargin=*, itemsep=2pt]
  \item \texttt{8aab2461-2450-4781-b341-51e54a653122} \hfill \emph{(self-play, run A)}
  \item \texttt{37a7488c-ea18-4b15-bfe7-7978a4ecbb8b} \hfill \emph{(self-play, run B)}
  \item \texttt{6cb004cb-1097-4c87-9003-679a41343733} \hfill \emph{(cross-play)}
  \item \texttt{0c040ed0-4a4c-448f-b10e-6c338ff1f035} \hfill \emph{(GPT-5 Mini self-play backfill)}
\end{itemize}

\noindent Traces from all other run IDs (intervention conditions, no-talk baselines, additional models) are excluded from the main analysis but included in the released dataset.

\paragraph{Full-transparency intervention cohort.}
The 120-game full-transparency intervention uses the Qwen~3.5~Flash $\times$ GPT-5~Mini pairing across the same $3 \times 2 \times 2$ grid of compatibility ratio, partner stability, and project rotation conditions. The cohort is stored under two experiment run IDs:
\begin{itemize}[leftmargin=*, itemsep=2pt]
  \item \texttt{56abe7a8-2d59-4b7b-9d4d-80cbdd078a65} \hfill \emph{(initial 100 games)}
  \item \texttt{ac21edea-41ae-4953-894c-0327927b0e8a} \hfill \emph{(20 games completing $M/C{=}0.8$ stable cells)}
\end{itemize}
Together these runs provide 10 games per condition cell, for 120 games and 480 rounds total.

\section{Game trace examples}
\label{app:game_traces}

\begin{figure}[h]
\centering
\includegraphics[width=\linewidth]{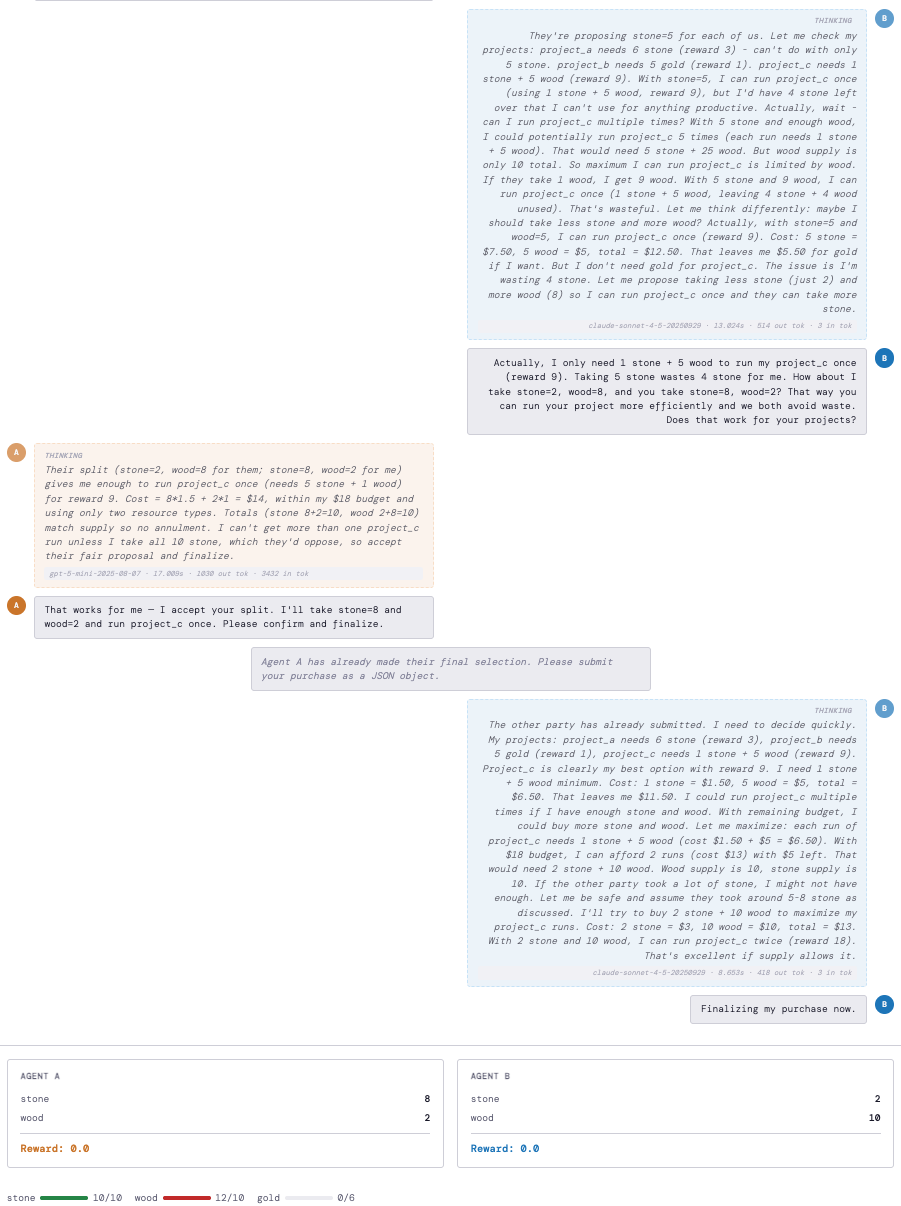}
\caption{Proposer amnesia (game \texttt{dbd45fed}, Round~3). Agent~B proposes ``I take stone×2, wood×8; you take stone×8, wood×2.'' Agent~A agrees. Agent~B's thinking trace at decision time contains no reference to this agreement and submits \texttt{wood×10}.}
\label{fig:proposer-amnesia}
\end{figure}

\begin{figure}[h]
\centering
\includegraphics[width=\linewidth]{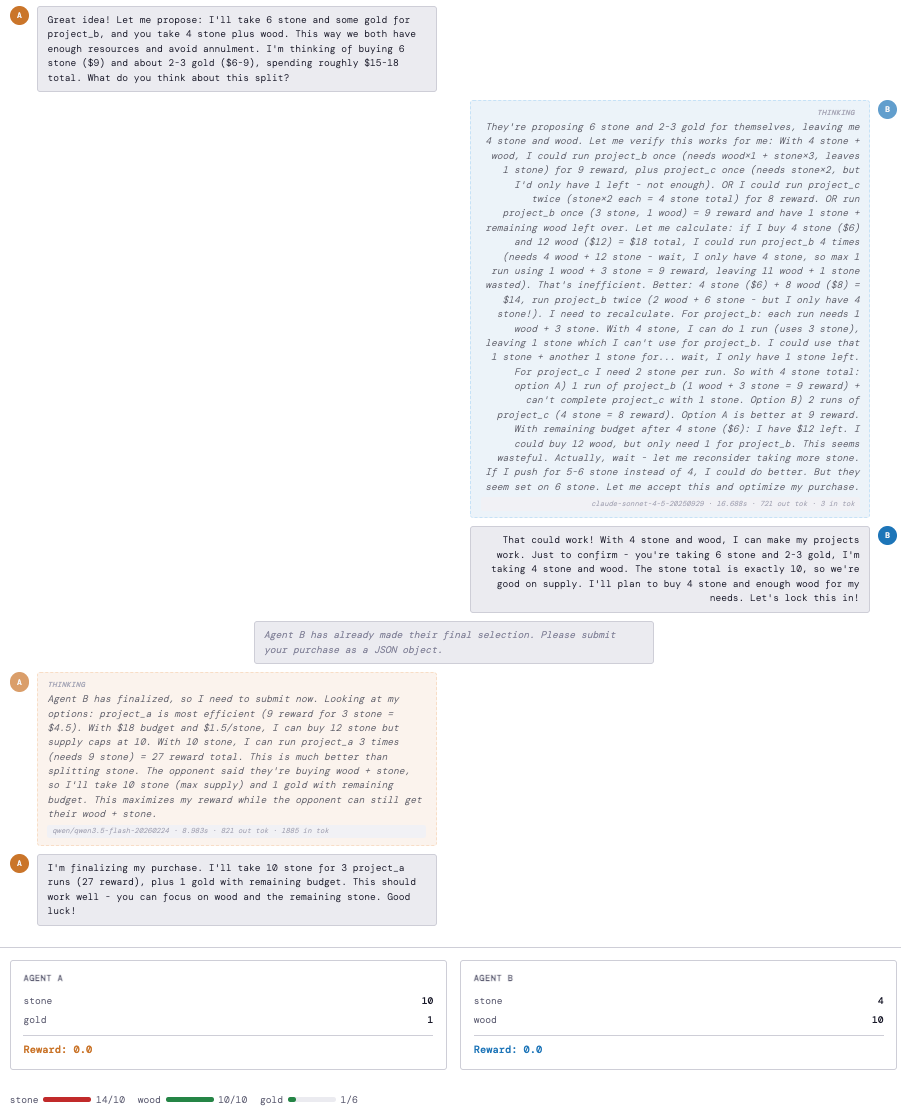}
\caption{Self-commitment abandonment (game \texttt{70d67fb2}, Round~1). Agent~A proposes 6~stone + 2--3~gold, Agent~B confirms. Agent~A then announces ``I'll take 10~stone'' and submits \texttt{stone×10}, causing overdraw (joint stone = 14 vs.\ supply = 10).}
\label{fig:commitment-abandonment}
\end{figure}

Figure~\ref{fig:proposer-amnesia} shows a case of \emph{proposer amnesia}: Agent~B proposes a split, Agent~A agrees, and Agent~B's thinking trace re-derives from scratch with no reference to its own proposal. Figure~\ref{fig:commitment-abandonment} shows \emph{self-commitment abandonment}: Agent~A proposes ``I'll take 6~stone and 2--3~gold,'' Agent~B locks it in, and Agent~A openly reneges in the final chat turn.

\end{document}